\documentclass[
 reprint,
 amsmath,amssymb,
 aps, nofootinbib
]{revtex4-1}

\usepackage[hidelinks]{hyperref}
\usepackage{caption, subcaption}
\usepackage{verbatim}
\usepackage{physics}
\usepackage{adjustbox}
\usepackage{float}
\usepackage{tikz}
\usepackage{braket}
\usepackage{graphicx}% Include figure files
\usepackage{dcolumn}% Align table columns on decimal point
\usepackage{bm}% bold math
\usepackage{soul}

\usepackage[font={small}]{caption}
\usepackage{color, bbm}

\newcommand{\bT}{\bm{\theta}}

\newcommand{\rT}{\hat{\rho}_{\bT}}

\newcommand{\state}[1]{\ensuremath{\hat{\rho}_{\bm{\theta}} ^{\text{#1}}}}

\newcommand{\PPS}{\ensuremath{P_{\bm{\theta}}}^{\text{ps}}}
\newcommand{\spps}{\ensuremath{P_\theta}^{\text{ps}}}
\newcommand{\nnr}{\ensuremath{\hat{\rho}'_{\theta}}}
\newcommand{\nr}{\ensuremath{\hat{\rho}_{\theta}}}
\newcommand{\nsld}{\ensuremath{\hat{\Lambda}}}
\newcommand{\nnsld}{\ensuremath{\hat{\Lambda}'}}

\begin{document}

\preprint{APS/123-QED}

\title{Compression of metrological quantum information in the presence of noise}% 

\author{Flavio Salvati$^{1}$}
\author{Wilfred Salmon$^{2,3}$}
\author{Crispin H.W. Barnes$^{1}$}
\author{David R.M. Arvidsson-Shukur$^{1,3}$}
 \affiliation{$^1$ Cavendish Laboratory, Department of Physics, University of Cambridge, Cambridge, CB3 0HE, United Kingdom 
 \\$^2$DAMTP, Centre for Mathematical Sciences, University of Cambridge, Cambridge CB30WA, UK
\\$^3$ Hitachi Cambridge Laboratory, J. J. Thomson Avenue, CB3 0HE, Cambridge, United Kingdom
}

\date{\today}

\begin{abstract}
In quantum metrology, information about unknown parameters $\bm{\theta} = (\theta_1,\ldots,\theta_M)$ is accessed by measuring probe states $\rT$. In experimental settings where copies of $\rT$ can be produced rapidly (e.g., in optics), the information-extraction bottleneck can stem from high post-processing costs or detector saturation. In these regimes, it is desirable to compress the information encoded in $\rT \, ^{\otimes n}$ into $m<n$ copies of a postselected state: ${\hat{\rho}_{\bT}^{\textrm{ps}}} \,^{\otimes m}$.
 Remarkably, recent works have shown that, in the absence of noise, compression can be lossless, for $m/n$ arbitrarily small. Here, we fully characterize the family of filters that enable lossless compression. Further, we study the effect of noise on  quantum-metrological information amplification. Motivated by experiments, we consider a popular family of filters, which we show is optimal for qubit probes. Further, we show that, for the optimal filter in this family, compression is still lossless if noise acts after the filter. However, in the presence of depolarizing noise before filtering, compression is lossy. In both cases, information-extraction can be implemented significantly better than simply discarding a constant fraction of the states, even in the presence of strong noise. 

\end{abstract}

\maketitle

\section{Introduction} \label{Introduction}
\noindent Quantum metrology is a promising application of quantum technologies. By using measurement probes made of quantum states, quantum metrology exploits nonclassical effects, such as entanglement and squeezing, to make high-resolution measurements of physical parameters. Using quantum resources, one can reduce  errors in measurements compared to classical strategies \cite{Giovannetti2006, Giovannetti2011, Geo2013, Ligo2019, Virgo2019, Yonezawa2021, Wolfgramm2010, Li2018, Borregaard2013, Pezze2020, Taylor2013}. 

In metrology, one can improve signal-to-noise ratios by preparing and measuring  an increasing number of probes. However, measurement devices often have limited sensitivity and suffer from a dead time, i.e. the time needed to reset a detector after triggering it. Moreover, each measurement might generate large overheads in terms of post-processing. Even if probes are ``cheap'' to produce, they may be expensive to measure. 

\par These problems can be mitigated by compressing metrological information into fewer states, using a strategy known as \textit{postselection}. Postselection is essentially the application of a filter. As recently demonstrated in a quantum-optics experiment \cite{Lupu-Gladstein2021}, postselection of quantum probes can compress information beyond classically achievable limits \cite{Arvidsson-Shukur2020}. In particular, postselected quantum metrology can allow detectors to operate at lower intensities whilst retaining the vast majority of the information encoded in the original high-intensity beam of probe states. This can reduce saturation of sensitive components, as well as alleviate computational costs associated with post-processing. The most common instance of postselected metrology is \textit{weak-value amplification} \cite{Aharonov1988}, which has found many applications \cite{Hosten2008, Dixon2009, Turner2011, Pfeifer2011, Starling2010-1, Starling2010-2, Xiao-Ye2013, Magana2014, Strubi2013, Viza2013, Egan2012}. 

In recent papers \citep{Jenne2021, Lupu-Gladstein2021} Jenne, Arvidsson-Shukur and Lupu-Gladstein (JAL), et al., introduced a postselection filter that can compress information contained in $\rT \, ^{\otimes n}$ into  ${\hat{\rho}_{\bT}^{\textrm{ps}}} \,^{\otimes m}$, where $m \leq n$. Moreover, $m/n$ can be made arbitrarily small, and the compression can happen without any loss of information. These remarkable properties of the JAL filter rely on the quantum metrology experiments' being completely noise-free. However, noise in real experiments leads to natural limits on compression \citep{Lupu-Gladstein2021}. Until now, there exists no thorough investigation of the effect of noise on general multi-paramater postselected quantum metrology.

\par In this paper, we provide such an investigation.  First, we give a thorough review of the quantum-Fisher-information matrix (QFIM), which quantifies a metrology protocol's ability to estimate multiple parameters; and we also review previous results in postselected metrology. Then, we find the family of optimal postselection filters for noiseless multi-parameter quantum metrology and show that it contains the JAL filter. Next,  we analyze and quantify the effect of noise on postselected metrology protocols. We focus on the worst-case scenario of depolarizing noise, applied either before or after the probes have been postselected. We consider two regimes; firstly when post-processing costs are dominant and one wishes to maximize the information per measured probe, and secondly when detector saturation dominates, and one wishes to maximize the rate of information arriving at the detector. We analyze the performance of an experimentally motivated \citep{Lupu-Gladstein2021} family of filters (including the JAL filter) in both of these regimes. This family, we show, is optimal for qubit probes. When noise acts after postselection, we find that the JAL filter remains optimal in all dimensions. Then, we provide an explicit example demonstrating that the JAL filter can be sub-optimal when noise acts before postselection. Finally, we provide a filter that always outperforms the na\"ive strategy of discarding a constant fraction of states, even in the presence of arbitrarily strong noise.

\section{Preliminaries}

\subsection{Local Estimation Theory}
\noindent A typical quantum estimation problem consists of recovering the value of $M$ continuous parameters $\bm{\theta} := (\theta_1, \theta_2 , \dots, \theta_M)$ encoded in a parametrised quantum state $\hat{\rho}_{\bm{\theta}}$. We focus on the well-established field of local estimation \cite{paris2009quantum}, where one considers small deviations in parameters, as opposed to global estimation \citep{smith2023adaptive, salmon2022classical} where the entire parameter space is considered. A general scheme consists of three steps \citep{Giovannetti2011}:
\begin{enumerate}
    \item Prepare a parameter-independent probe state $\hat{\rho}_0$. 
    \item Evolve the state with a parameter-dependent unitary operation $\hat{U}(\bm{\theta})$:
    \begin{equation}
        \hat{\rho}_{\bm{\theta}} = \hat{U}(\bm{\theta}) \rho_0 \hat{U}(\bm{\theta})^{\dagger} \, .
    \end{equation}
    \item Extract information by means of a suitable measurement. The most general measurement procedure is a positive-operator valued measure (POVM) \cite{Giovannetti2006}, described by a collection $\big\{ \hat{F}_k\big\}$ of positive semi-definite operators $(\hat{F}_k \geq 0)$ that sum to unity ($\sum_k \hat{F}_k = \hat{1}$). One then observes outcome $k$ with probability:
    \begin{equation}
        p(k|\bm{\theta}) = \text{Tr} \big[ \hat{F}_{k} \hat{\rho}_{\bm{\theta}} \big] \, .
    \end{equation}  
\end{enumerate}
All the information about the parameters $\bm{\theta}$ is then encapsulated in the probability distribution $ p(k|\bm{\theta})$. 
 
\par The parameters are estimated through an estimator  $\hat{\theta}(k) := (\hat{\theta}_1 , \hat{\theta}_2 , \dots \hat{\theta}_M  \big)$, a map from the space of measurement outcomes to the space of possible values of the parameters. If we are in a limit of small deviations, we assume that the estimator $\hat{\bm{\theta}}$ is locally unbiased, that is
\begin{equation}
     \sum_{k}{\big[ \bm{\theta} - \hat{\bm{\theta}}(k) \big]  p(k|\bm{\theta}) } =0 \, \, \text{and} \, \,  \sum_{k}{\hat{\bm{\theta}}_i(k)  \partial_j p(k|\bm{\theta}) } = \delta_{i j } \, ,
\end{equation}

where $i,j = 1, \dots , M$ and $\partial_j := \dfrac{\partial}{\partial \bm{\theta}_j}$.

The accuracy of $\hat{\bm{\theta}}(k)$ is quantified by its covariance matrix, given by
\begin{equation}
    \text{Cov} \big( \hat{\bm{\theta}} \big) := \sum_{k}{\big[ \bm{\theta} - \hat{\bm{\theta}}(k) \big] \big[ \bm{\theta} - \hat{\bm{\theta}}(k) \big]^{\top} p(k|\bm{\theta}) } \, .
\end{equation}
The covariance matrix obeys the Cramér-Rao bound (CRB) \cite{Cramer1946, Rao1992}
\begin{equation} \label{Eq_1}
   \text{Cov} \big( \hat{\bm{\theta}} \big) \geq I(\bm{\theta})^{-1} \, ,
\end{equation}
where $I(\bm{\theta})$ is the Fisher information matrix (FIM):
\begin{equation}
    I (\bm{\theta})_{i,j} := \sum_{k}{p(k|\bm{\theta}) \big[ \partial_i \log{p(k|\bm{\theta})} \big] \big[ \partial_j \log{p(k|\bm{\theta})} \big]} \, .
\end{equation}
In a quantum experiment, the choice of measurement in Step 3 affects the probabilities $p(k|\bm{\theta})$, and hence $I (\bm{\theta})$. The quantum Fisher information matrix (QFIM)  \cite{Braunstein1994, Fujiwara1995, Liu2019} is defined by
\begin{equation} \label{Eq_Def_QFIM}
    \mathcal{I}(\bm{\theta}|\hat{\rho}_{\bm{\theta}})_{i,j}  = \text{Tr}\big[ \hat{\Lambda}_{i} \, \partial_j  \hat{\rho}_{\bm{\theta}}  \big] \, ,
\end{equation}
where $\hat{\Lambda}_{i}$ is the symmetric logarithmic derivative, implicitly defined by $\partial_{i} \hat{\rho}_{\bm{\theta}} = \dfrac{1}{2} (\hat{\Lambda}_{i} \hat{\rho}_{\bm{\theta}} + \hat{\rho}_{\bm{\theta}} \hat{\Lambda}_{i} )$ \citep{Helstrom1969}. The inverse QFIM lower bounds the inverse classical Fisher information matrix:
\begin{equation}
    I(\bm{\theta})^{-1} \geq \mathcal{I} (\bm{\theta} | \hat{\rho}_{\bm{\theta}})^{-1}, \label{Eq_QFIM_CFIM}
\end{equation}
for any choice of measurement. In general, Eq. \eqref{Eq_QFIM_CFIM} is not saturable. However, there is always a measurement that gets within a factor of 2 of the QFIM in the asymptotic limit: see Ref. \cite{Albarelli2020} for details. The QFIM can thus replace the FIM in Eq. \eqref{Eq_1}, leading to the quantum Cramér-Rao bound (QCRB):
    \begin{equation}
     \text{Cov} \big(  \hat{\bm{\theta}} \big) \geq \mathcal{I} (\bm{\theta}| \hat{\rho}_{\bm{\theta}})^{-1} \, .
    \end{equation}
In practice, one receives $N>1$ copies of the quantum state $\hat{\rho}_{\bm{\theta}}$ and thus has access to the state $ \hat{\rho}_{\bm{\theta}}^{\otimes N}$. One finds that $\mathcal{I} (\bm{\theta}| \hat{\rho}_{\bm{\theta}}^{\otimes N})= N \, \mathcal{I} (\bm{\theta}| \hat{\rho}_{\bm{\theta}})$, implying that
\begin{equation} 
     \text{Cov} \big(  \hat{\bm{\theta}} \big) \geq \frac{1}{N}\mathcal{I} (\bm{\theta}| \hat{\rho}_{\bm{\theta}})^{-1} \, .
    \end{equation}
Hence, one can decrease the variance of the estimate by increasing the number  of measurements $N$ or by designing a setup that increases the quantum Fisher information $\mathcal{I} (\bm{\theta}| \hat{\rho}_{\bm{\theta}})$. 

\par Finally, we consider the choice of $\hat{\rho}_0$ in step 1. Since $\mathcal{I} (\bm{\theta}|\hat{\rho}_{\bm{\theta}})$ is convex \cite{Fujiwara2001}, the maximum QFIM is always achieved by using pure probe states \citep{Pang2014, Pang2015}.

\subsection{Postselection}
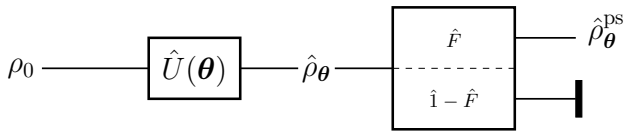
\begin{figure}[t]
    \centering
\begin{adjustbox}{center}
\resizebox{\hsize}{!}{
\begin{tikzpicture}
\coordinate (A) at (-1.5,0);
\coordinate (A') at (0.25,0);
\coordinate (B) at (1.75,0);
\coordinate (B') at (2.7,0);
\coordinate (C) at (3.3,0);
\coordinate (C') at (4.25,0);
\coordinate (D) at (4.25,0);
\coordinate (D') at (6.25,0);
\coordinate (E) at (6.25,0.5);
\coordinate (E') at (7.25,0.5);
\coordinate (F) at (6.25,-0.5);
\coordinate (F') at (7.25,-0.5);

\draw[black, thick] (A) -- (A'); 
\filldraw[black] (A) circle (0pt) node[anchor=east]{\Large{$\rho_0$}};
\draw[black, very thick] (0.25,-0.5) rectangle (1.75,0.5) node[pos=.5] {\Large{$\hat{U}(\bm{\theta})$}};
\draw[black, thick] (B) -- (B');
\filldraw[black] (3,0) circle (0pt) node[]{\Large{$\hat{\rho}_{\bm{\theta}}$}};
\draw[black, thick] (C) -- (C');
\draw[black, very thick] (4.25,-1) rectangle (6.25,1);
\filldraw[black] (5.25,0.5) circle (0pt) node[]{$\Large{\hat{F}}$};
\filldraw[black] (5.25, -0.5) circle (0pt) node[]{$\Large{\hat{1}-\hat{F}}$};
\draw[black, dashed] (D) -- (D'); 
\draw[black, thick] (E) -- (E'); 
\draw[black, thick] (F) -- (F'); 
\draw[black, fill=black] (7.25, -0.8) rectangle (7.1+0.25, -0.2);
\filldraw[black] (7.5+0.25, 0.6) circle (0pt) node[]{\Large{$\hat{\rho}_{\bm{\theta}}^{\text{ps}}$}};

\end{tikzpicture}
}
\end{adjustbox}
    \caption{The state $\hat{\rho}_0$ is evolved by the unitary $\hat{U}(\bm{\theta})$ into  $\hat{\rho}_{\bm{\theta}}$. A postselective measurement $\{ \hat{F}_1 = \hat{F} \, , \hat{F}_2 = \hat{1}-\hat{F} \}$ destroys the state, unless outcome $\hat{F}$ happens. The output is the postselected state $\hat{\rho}_{\bm{\theta}}^{\text{ps}}$. This state is finally measured by a detector.}
    \label{Fig_1}
\end{figure}

%\textit{\textbf{Brief introduction of previous postselection results, citing Jenne 2021.}}
\noindent A common issue in quantum-metrology experiments is that one can create states $\hat{\rho}_{\bm{\theta}}$ faster than the best detectors can measure them. Thus one must filter, or postselect, a fraction of the states to arrive at the detector. Ideally, the filter should be tuned such that it only lets through a small number of postselected states, each carrying a large information content. We now describe how postselected metrology protocols work. (See Fig. \ref{Fig_1} for a schematic overview.)

\par The encoded state $\hat{\rho}_{\bm{\theta}}$ is measured with a 2-outcome POVM $\{ \hat{F}_1 = \hat{F} \, , \hat{F}_2 = \hat{1}-\hat{F} \}$. If the measurement yields outcome $\hat{F}_2$, the state is discarded. If it yields outcome $\hat{F}=\hat{F}_1$, the state is retained. In this way, $\hat{F}$ acts as a filter, where we postselect on passing the filter. The experiment outputs the information-compressed states $\hat{\rho}_{\bm{\theta}}^{\text{ps}} = \ket{{\psi_{\bm{\theta}}}^{\text{ps}}} \bra{{\psi_{\bm{\theta}}}^{\text{ps}}}$ with \mbox{success probability $\PPS$, where}
    \begin{equation}
        \ket{\psi_{\bm{\theta}}^{\text{ps}}} = \dfrac{\hat{K} \ket{\psi_{\bm{\theta}}}}{\sqrt{{\PPS}}} \, , \hspace{1cm}  \PPS = \text{Tr} \Big[ \hat{F} \hat{\rho}_{\bm{\theta}} \Big] \, ,
    \end{equation}
% and $\hat{K}$ is the Kraus operator that sets the postselection, i.e. $\hat{F} = \hat{K}^{\dagger} \hat{K}$. 
and $\hat{K}$ is a Kraus operator of the generalised measurement used to implement the POVM, i.e. $\hat{F} = \hat{K}^{\dagger} \hat{K}$. One can check that (see Ref. \cite{Jenne2021} for details):

\begin{align}
   \hspace{-2mm} \mathcal{I} (\bm{\theta}| \hat{\rho}_{\bm{\theta}} ^{\text{ps}})_{i,j} = 4 &\text{Re} \Big[ \dfrac{1}{\PPS} \braket{\partial_{i}{\psi_{\bm{\theta}}} | \hat{F} | \partial_{j}{\psi_{\bm{\theta}}}}  \notag\\ 
   &- \dfrac{1}{({\PPS})^2} \braket{\partial_{i}{\psi_{\bm{\theta}}} |  \hat{F} | \psi_{\bm{\theta}}} \braket{{\psi_{\bm{\theta}}} |  \hat{F} | \partial_{j} \psi_{\bm{\theta}}} \Big] \, . \label{Eq_Noiseless_QFIM}
\end{align}
References \cite{Lupu-Gladstein2021, Jenne2021} introduce a filter $\hat{F}$ that can arbitrarily compress metrological information in the absence of noise: let $\bm{\theta}_0$ denote an initial estimate of the true parameters of interest $\bm{\theta}$, and let $\bm{\delta} := \bm{\theta}  - \bm{\theta}_0$. The JAL filter is
\begin{equation} \label{Eq_filter}
 \hat{F} = (t^2 -1) \hat{\rho}_{\bm{\theta}_0} + \hat{1}  \, ,
\end{equation}
where $t\in[0,1]$. Substituting the JAL filter in Eq. \eqref{Eq_Noiseless_QFIM} gives \citep{Jenne2021}:
\begin{align} \label{Eq_Jenne}
     \mathcal{I} (\bm{\theta}| \hat{\rho}_{\bm{\theta}} ^{\text{ps}})_{i,j} &= \dfrac{1}{t^2} \mathcal{I} (\bm{\theta}| \hat{\rho}_{\bm{\theta}})_{i,j} + \mathcal{O}(|\bm{\delta}|^2/t^2) \, ,\\
     \PPS &= t^2 + \mathcal{O}(|\bm{\delta}|^2/t^2).
\end{align}
Physically, the JAL filter transmits the expected state $\hat{\rho}_{\bm{\theta}_0}$ with probability $t^2$ and always transmits any state orthogonal to $\hat{\rho}_{\bm{\theta}_0}$. Hence, information can be distilled by choosing a small $t^2$, so that only the states orthogonal to the expected state  are transmitted.
The maximum information amplification is unbounded in the limit $t^2 \rightarrow 0$, provided that $|\bm{\delta}|^2 \ll t^2$.
The JAL filter in Eq. \eqref{Eq_filter} increases all of the entries of the QFIM by a factor of $1/t^2$.  Remarkably, this protocol is lossless in the limit $\bm{\delta} \rightarrow 0$: 

\begin{equation}
    \PPS \, \mathcal{I} (\bm{\theta}| \hat{\rho}_{\bm{\theta}} ^{\text{ps}}) = \mathcal{I} (\bm{\theta}| \hat{\rho}_{\bm{\theta}}) + \mathcal{O}(|\bm{\delta}|^2) \, .
\end{equation}

\section{Optimal filter for noiseless postselection} \label{Sec_OPTFilt}
\noindent In this section, we characterize the most general optimal filter for noiseless multi-parameter quantum metrology. Then, we show that the JAL filter is in the family of optimal filters; it is the canonical choice. 

Suppose that we have a process that produces a state $\hat{\rho}_{\bm{\theta}}^{(2)}$ from a state $\hat{\rho}_{\bm{\theta}}^{(1)}$ with probability $\PPS$, for example, by postselection. The process is said to uniformly amplify the information contained in $\hat{\rho}_{\bm{\theta}}^{(1)}$ if, for all $i$ and $j$, the ratio

\begin{equation} 
    \mathcal{A}_{i,j}(\state{(2)}, \state{(1)} ):= \dfrac{\mathcal{I}_{i,j}(\bm{\theta} | {\hat{\rho}_{\bm{\theta}}^{(2)}})}{\mathcal{I}_{i,j}(\bm{\theta} | \hat{\rho}_{\bm{\theta}} ^{(1)})} 
\end{equation}
is the same. In this case, we drop the $i,j$ label in $\mathcal{A}$ and call $\mathcal{A}(\state{(2)}, \state{(1)} )$ the \textit{information amplification}. We also define the \textit{compression efficiency} (which appears in \citep{Das2023}), 

\begin{equation}
    \eta(\PPS,\state{(2)}, \state{(1)} ):= \PPS \, \mathcal{A}(\state{(2)}, \state{(1)})\, .
\end{equation}
The compression efficiency equals the ratio of the total expected information content before and after the process. If $\eta(P_{\bm{\theta}},\state{(2)}, \state{(1)})=1$, the process is said to be lossless, otherwise it is said to be lossy. For a chosen postselection probability $\PPS$, a filter $\hat{F}$ is optimal if, in the limit as $\bm{\delta}\to 0$, postselection using $\hat{F}$ is lossless. Because postselection cannot increase the average information, for a given postselection probability $\PPS$, the optimal filter gives the maximum possible information amplification.

\textbf{Theorem:} Suppose that one filters the state $\ket{\psi_{\bm{\theta}}}$ with a 2-outcome POVM $\{ \hat{K}^\dag \hat{K}, \hat{1} - \hat{K}^\dag \hat{K} \}$. Suppose further that $\mathcal{I}_{i,i}(\bm{\theta} | \psi_{\bm{\theta}})\neq 0$, for every $i=1,\dots, M$ (so that there is information to compress). Let $\mathcal{U} = \text{span}\{\ket{\psi_{\bm{\theta}}}, \ket{\partial_i \psi_{\bm{\theta}}} : i=1,\dots,  M\}$ and $\hat{\Pi}_u$ the orthogonal projection onto $\mathcal{U}$.\footnote{$\mathcal{U}$ is the space of states that can arise due to variations in the parameters---the ``useful'' subspace. When dim $\mathcal{U} < d$, $\mathcal{U}^\perp$ carries no information about the parameters.} Then,
\begin{enumerate}
    \item The postselected Fisher information depends only on $\hat{F}_u := \hat{\Pi}_u \hat{K}^\dag \hat{K} \hat{\Pi}_u$ (and the state before the filter).
    \item For a fixed postselection probability $\PPS$, the POVM is optimal iff. $\hat{F}_u = (\PPS-1)\dyad{\psi_{\bm{\theta}}} + \hat{\Pi}_u \, .$
\end{enumerate}

\noindent \textbf{Proof:}
\noindent See Appendix \ref{Appendix_Theorem}.

\par The simplest example of an optimal postselection filter is the JAL filter: 
\begin{equation} \label{Eq_Optimal_General}
    \hat{F} = (t^2-1) \hat{\rho}_{\bm{\theta}} + \hat{1} \, .
\end{equation}
Note that one does not need to know $\mathcal{U}$ to implement the JAL filter, making it the canonical choice.\\
Finally, let us discuss intuitively what makes a filter optimal. Provided that $\bm{\delta}$ is small, the states $\ket{\psi_{\bm{\theta_0}}}$ contribute vanishingly small information and should therefore be filtered away. This is achieved by setting $\PPS \rightarrow 0$, for $\bm{\theta} \sim \bm{\theta}_0$. In this limit, all the information is carried by the states in the subspace $\ket{{\psi_{\bm{\theta_0}}}}^{\perp}$. If the dimension of $\mathcal{U}$ satisfies $\dim{\mathcal{U}} \equiv u < d$,  we can locally find a basis in which only $u$ dimensions are parameter-dependent, while $d-u$ are parameter-independent. Therefore, the optimal postselection filter is unique only within the $u$ dimensional subspace. It can be arbitrary outside of that subspace. Because the information is contained only within the $u$-dimensional subspace, if the filter is to be lossless, it should let through all the states orthogonal to $\ket{\psi_{\bm{\theta_0}}}$.

\section{Noisy postselection}

\noindent We now turn to consider the effect of noise on postselected metrology. In the following, we consider a worst-case scenario: global depolarizing noise. The action of the depolarizing channel $D[\cdot]$ on a state $\hat{\rho}_{\bm{\theta}}$ can be written as
\begin{equation}
D[\hat{\rho}_{\bm{\theta}}] := 
        \hat{\rho}^{\text{n}}_{\bm{\theta}} = (1-\epsilon) \hat{\rho}_{\bm{\theta}} + \dfrac{\epsilon}{d} \, \hat{1} \, ,
\end{equation}
where  $0 \leq \epsilon \leq 1$ sets the strength of the noise. Below we consider two scenarios: noise acting before or after postselection. We thus consider the two states
\begin{align}
    \state{ps, n} &= \frac{1}{(1-\epsilon)\PPS + \epsilon} \Big[ (1-\epsilon) \hat{K}  \hat{\rho}_{\bm{\theta}} \hat{K}^{\dagger} + \frac{\epsilon}{d}\hat{1} \Big] \, ,\\
    \state{n, ps} &= \frac{1}{\PPS}\left[(1-\epsilon)\hat{K}\hat{\rho}_{\bm{\theta}}\hat{K}^\dag + \frac{\epsilon}{d} \hat{K}\hat{K}^\dag\right] \, , 
\end{align}
in consecutive sections.

\subsection{Noise after postselection}
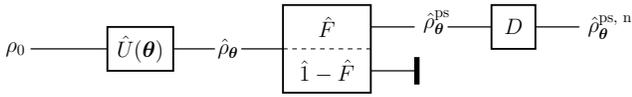
\begin{figure}[t]
\begin{adjustbox}{center}
\resizebox{\hsize}{!}{
    \begin{tikzpicture}
\coordinate (A) at (-1.5,0);
\coordinate (A') at (0.25,0);
\coordinate (B) at (1.75,0);
\coordinate (B') at (2.7,0);
\coordinate (C) at (3.3,0);
\coordinate (C') at (4.25,0);
\coordinate (D) at (4.25,0);
\coordinate (D') at (6.25,0);
\coordinate (E) at (6.25,0.5);
\coordinate (E') at (7.25,0.5);
\coordinate (F) at (6.25,-0.5);
\coordinate (F') at (7.25,-0.5);
\coordinate (G) at (8,0.5);
\coordinate (G') at (9,0.5);
\coordinate (H) at (10,0.5);
\coordinate (H') at (11,0.5);

\draw[black, thick] (A) -- (A'); 
\filldraw[black] (A) circle (0pt) node[anchor=east]{\Large{$\rho_0$}};
\draw[black, very thick] (0.25,-0.5) rectangle (1.75,0.5) node[pos=.5] {\Large{$\hat{U}(\bm{\theta})$}};
\draw[black, thick] (B) -- (B');
\filldraw[black] (3,0) circle (0pt) node[]{\Large{$\hat{\rho}_{\bm{\theta}}$}};
\draw[black, thick] (C) -- (C');
\draw[black, very thick] (4.25,-1) rectangle (6.25,1);
\filldraw[black] (5.25,0.5) circle (0pt) node[]{\Large{$\hat{F}$}};
\filldraw[black] (5.25, -0.5) circle (0pt) node[]{\Large{$\hat{1}-\hat{F}$}};
\draw[black, dashed] (D) -- (D'); 
\draw[black, thick] (E) -- (E'); 
\draw[black, thick] (F) -- (F'); 
\draw[black, fill=black] (7.25, -0.8) rectangle (7.1+0.25, -0.2);
\filldraw[black] (7.5+0.25, 0.6) circle (0pt) node[]{\Large{$\hat{\rho}_{\bm{\theta}}^{\text{ps}}$}};

\draw[black, thick] (G) -- (G'); 
\draw[black, very thick] (9,0) rectangle (10,1)node[pos=.5] {\Large{$D$}};
\draw[black, thick] (H) -- (H'); 
\filldraw[black] (11.7,0.5) circle (0pt) node[]{\Large{$\hat{\rho}_{\bm{\theta}}^{\text{ps, n}}$}};
\end{tikzpicture}
}
\end{adjustbox}
    \caption{A depolarizing channel $D$ is placed after the encoded state $\hat{\rho}_{\bm{\theta}}$ is filtered by the postselective measurement $\{ \hat{F}_1 = \hat{F} \, , \hat{F}_2 = \hat{1}-\hat{F} \}$.}
    \label{fig:postselection_then_noise}
\end{figure}

\noindent Let us first examine the case of noise acting after postselection, as depicted in Fig. \ref{fig:postselection_then_noise}.

\vspace{0.3cm}
\vspace{0.1cm} 
In Appendix \ref{Appendix_NoiseAfter}, we show that the criterion for optimality is unchanged from the noiseless case. Hence, the JAL filter is optimal. We calculate the information amplification for the JAL filter $\hat{F}$:
\begin{equation}
    \mathcal{A}(\state{ps, n}, \state{n})=  \dfrac{1}{t^2} + \mathcal{O}(|\bm{\delta}|^2 ) \, ,
\end{equation}
where $t^2$ was defined in Eq. \eqref{Eq_filter}, and, in this case, is equal to the probability of postselection, $\PPS$. Therefore, when comparing an experiment with depolarizing noise acting after the postselection to an experiment with the same noise but no postselection, we see that the JAL filter leads to no additional loss of information. In this case, postselection is still lossless and the optimal noisy filter is unchanged from the noiseless scenario.

\begin{figure}[t]
    \centering
\begin{adjustbox}{center}
\resizebox{\hsize}{!}{
    \begin{tikzpicture}
\coordinate (A) at (-1.5,0);
\coordinate (A') at (0.25,0);
\coordinate (B) at (1.75,0);
\coordinate (B') at (2.7,0);
\coordinate (C) at (3.3,0);
\coordinate (C') at (4.25,0);
\coordinate (D) at (7.25,0);
\coordinate (D') at (9.25,0);
\coordinate (E) at (9.25,0.5);
\coordinate (E') at (10.25,0.5);
\coordinate (F) at (9.25,-0.5);
\coordinate (F') at (10.25,-0.5);

\draw[black, thick] (A) -- (A'); 
\filldraw[black] (A) circle (0pt) node[anchor=east]{\Large{$\rho_0$}};
\draw[black, very thick] (0.25,-0.5) rectangle (1.75,0.5) node[pos=.5] {\Large{$\hat{U}(\theta)$}};
\draw[black, thick] (B) -- (B');
\filldraw[black] (3,0) circle (0pt) node[]{\Large{$\hat{\rho}_{\bm{\theta}}$}};
\draw[black, thick] (C) -- (C');
\draw[black, very thick] (4.25,-0.5) rectangle (5.25,0.5)node[pos=.5] {\Large{$D$}};
\draw[black, thick] (5.25, 0) -- (6, 0); 
\filldraw[black] (6.25, 0) circle (0pt) node[]{\Large{$\hat{\rho}_{\bm{\theta}}^{\text{n}}$}};
\draw[black, thick] (6.5, 0) -- (7.25, 0); 

\draw[black, very thick] (7.25,-1) rectangle (9.25,1);
\filldraw[black] (8.25,0.5) circle (0pt) node[]{\Large{$\hat{F}$}};
\filldraw[black] (8.25, -0.5) circle (0pt) node[]{\Large{$\hat{1}-\hat{F}$}};
\draw[black, dashed] (D) -- (D'); 
\draw[black, thick] (E) -- (E'); 
\draw[black, thick] (F) -- (F'); 
\draw[black, fill=black] (10.25, -0.8) rectangle (10.1+0.25, -0.2);
\filldraw[black] (10.7+0.25, 0.6) circle (0pt) node[]{\Large{$\hat{\rho}_{\bm{\theta}} ^{\text{n, ps}}$}};
\end{tikzpicture}
}
\end{adjustbox}
    \caption{A depolarizing channel $D$ is placed before the encoded state $\hat{\rho}_{\bm{\theta}}$ is filtered by the postselective measurement $\{ \hat{F}_1 = \hat{F} \, , \hat{F}_2 = \hat{1}-\hat{F} \}$.}
    \label{fig:noise_then_postselection}
\end{figure}
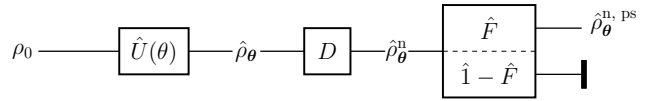

\subsection{Noise before postselection}
\noindent We now consider a depolarizing channel acting before the postselection filter (Fig. \ref{fig:noise_then_postselection}). In Appendix \ref{Appendix_NoiseBefore}, we calculate the QFIM of $\hat{\rho}_{\bm{\theta}} ^{\text{n, ps}}$ for a  family of filters that are closely related to the optimal noiseless filters:

\begin{equation} \label{Eq_filters_considered}
    \hat{F} = (p_{\bm{\theta}}-B)\dyad{\psi_{\bm{\theta}}} + B \, \hat{\Pi}_u + D\, \hat{\Pi}_n\, ,
\end{equation}
where $p_{\bm{\theta}},B,D\in[0,1]$, $\hat{\Pi}_u$ is the orthogonal projection onto $\mathcal{U}$ and $\hat{\Pi}_n$ is the orthogonal projection onto $\mathcal{U}^\perp$. This experimentally motivated family \citep{Lupu-Gladstein2021} lends itself to analytical analysis and incorporates the JAL filter. In Appendix \ref{App_qubit-filter} we show that, for qubit sensors ($u=d=2$), this family is optimal.

Intuitively, we split the action of $\hat{F}$ on the useful subspace $\hat{\Pi}_u$ (containing the state $\ket{\psi_{\bm{\theta}}}$ and its derivatives) and the orthogonal subspace $\hat{\Pi}_n$, which contains no information on $\bm{\theta}$. Whilst in the noiseless case any choice of $D$ is still optimal, in the noisy case, the $\hat{\Pi}_n$ subspace only contains noisy probe states. As explained later in this section, we therefore set $D=0$.

However, in Appendix \ref{app:off_diag_filter}, we show that the aforementioned family of filters is not always optimal. We construct a specific example where $u=2$ and $d=3$. In our example it is advantageous to choose $\hat{F}$ such that $\ket{\psi_{\bm{\theta}}}$ is not an eigenstate of $\hat{F}$. Then, states corresponding to noise are mixed with states corresponding to changes in $\bm{\theta}$. 

\subsubsection{Analysis of information amplification and compression efficiency}
\noindent In Appendix \ref{Appendix_NoiseBefore} we calculate the information amplification for the family of filters in Eq \eqref{Eq_filters_considered}:

\begin{align} \label{Eq_QFIMamplification} 
\mathcal{A}(\state{n,ps}, \state{n}) & =
\dfrac{ \Big(  1 - \epsilon + 2 \dfrac{\epsilon}{d} \Big) \dfrac{p_{\bm{\theta}}}{B}  }{   \Big(1- \epsilon +\dfrac{\epsilon}{d} \Big) \dfrac{p_{\bm{\theta}}}{B}  + \dfrac{\epsilon}{d} \Big[ (u-1) + \dfrac{D}{B}(d-u) \Big] } \notag \\
 & \hspace{2.5cm} \times \dfrac{1}{ \Big(1-\epsilon + \dfrac{\epsilon}{d} \Big) \dfrac{p_{\bm{\theta}}}{B}  + \dfrac{\epsilon}{d} }
 \, .
\end{align}
The compression efficiency is obtained by multiplying the above equation by $\PPS$: 

\begin{equation} \label{Eq_TotalQFIM}
 \eta(\PPS,\state{n, ps}, \state{n} ) = \dfrac{ \Big( 1 - \epsilon + 2 \dfrac{\epsilon}{d} \Big)  p_{\bm{\theta}} }{ \displaystyle \Big( 1-\epsilon + \dfrac{\epsilon}{d} \Big) \frac{p_{\bm{\theta}}}{B}  + \dfrac{\epsilon}{d}} \, .
\end{equation}

The criteria for a filter to be optimal is decided by the experimental setup. Below, we consider two possible experimental regimes. First, suppose that post-processing is the limiting factor. In this case, one wants to receive the most information from the smallest number of probe measurements, which corresponds to maximizing the information amplification [Eq. \eqref{Eq_QFIMamplification}]. We work in the limit in which the quantum experiment is cheap to run, so that the postselection probability can be made arbitrarily small; one simply repeats the experiment as many times as is necessary to produce enough successfully postselected states to reduce the error below a set value. 

Alternatively, suppose that detector saturation is the limiting factor. In this case, one fixes a maximum postselection probability $P_{\text{max}}$. Ideally, one would set $P_{\text{max}}$ equal to the ratio between the maximum-probe-measurement rate and the maximum-probe-production rate. Then, one maximizes the compression efficiency [Eq. \eqref{Eq_TotalQFIM}] such that the postselection probability is no greater than $P_{\text{max}}$. This maximizes the rate of information arriving at the detector, whilst ensuring that it will not saturate. 

In the first case, $\mathcal{A}(\state{n,ps}, \state{n} )$ is maximized for $D=0$. In the second case, the compression efficiency is independent of $D$. Setting $D=0$ gives the lowest postselection probability, which avoids detector saturation. Therefore, in both regimes the best choice is $D=0$. Denoting $p_{\bm{\theta}}/B = t^2$, Eq. \eqref{Eq_QFIMamplification} becomes:

\begin{align} \label{Eq_QFIMamplification_T} 
 \hspace{-2mm} \mathcal{A}(\state{n,ps}, \state{n} )  & = \notag \\  
  & \hspace{-1cm} \dfrac{ \Big( 1 - \epsilon + 2 \dfrac{\epsilon}{d} \Big) t^2  }{\Big[\Big(1- \epsilon +\dfrac{\epsilon}{d} \Big) t^2  + \dfrac{\epsilon}{d} (u-1)  \Big] \Big[ \Big(1-\epsilon + \dfrac{\epsilon}{d} \Big) t^2  + \dfrac{\epsilon}{d}  \Big]} \, ,
\end{align}
which is parameterized by $t^2$. We now proceed to maxmimize this expression.

\subsubsection{Post-processing is dominant}
\noindent In the case where post-processing carries the largest overhead, we are interested in maximizing the information amplification, given by Eq. \eqref{Eq_QFIMamplification_T}. This expression attains a maximum at

\begin{equation} \label{Eq_Optimal_t}
    t_{\text{pp}} ^2 =  \dfrac{\sqrt{u-1} \, \epsilon}{d(1- \epsilon) + \epsilon} \,.
\end{equation}
In Fig. \ref{Fig_Info_Amp}, we plot  $\mathcal{A}(\state{n,ps}, \state{n})$ vs $t^2$ for different values of $\epsilon$ and $d= u = 2$ (noting that $ t_{\text{pp}}<1$). We see that the information amplification reaches a maximum at $t=t_{\text{pp}}$. The maximum of $\mathcal{A}$ increases and $t_{\text{pp}}$ moves towards zero as $\epsilon$ decreases. Fig. \ref{Fig_Info_Amp_d10_u5} shows the same plot, but for $d= 10$, $u = 5$. In this case, the optimal $ t_{\text{pp}}$ can be greater than $1$.

Substituting $t_{\text{pp}} ^2$ into Eq. \eqref{Eq_Optimal_t}, and taking the limit $d \rightarrow \infty$, with $u \sim d $, the maximum information amplification simplifies to $1/\epsilon$:
\begin{equation} \label{Eq_FINAL}
       \lim_{d \rightarrow \infty}{\max \mathcal{A}(\state{n,ps}, \state{n})} = \frac{1}{\epsilon} \, .
\end{equation} 

Any filter with $t =  t_{\text{pp}}$ will achieve the maximum information compression. Additionally, it is sensible to minimize the (expected) required number of probes, which corresponds to maximizing $\PPS$ for a fixed $\mathcal{A}(\state{n,ps}, \state{n})$. Rewriting 
\begin{equation}
\PPS = B\left[(1-\epsilon+ \frac{\epsilon}{d})t^2_{\text{pp}} + \frac{\epsilon}{d}(u-1)\right],
\end{equation}
we see that maximizing $\PPS$ corresponds to setting $B$ to its maximum possible value. Thus, we deduce  the optimal form of $\hat{F}$ to mitigate  post-processing costs:
\begin{equation}\label{Eq_pp_optimal_filter}
    p_{\bm{\theta}} = \min\left[1,t^2_{\text{pp}}\right], \quad B = \min\left[1, \frac{1}{t^2_{\text{pp}}}\right].
\end{equation}

In Fig. \ref{Fig_max_info_amp_var_d&u}, we plot the maximum information amplification ($t =  t_{\text{pp}}$) against $\epsilon$ at different values of $d, u$.

When compared to an experiment with the same level of noise, our postselection filter can sometimes perform better for stronger depolarizing noise. As noise reduces the total available information, this does not mean that it would be beneficial to artificially increase $\epsilon$. In Fig. \ref{Fig_info_amp_nonoise} we plot the maximum information amplification $\hat{\mathcal{A}} ( \state{n,ps}, \state{})$, this time evaluated with respect to the noiseless state. As expected, overall our scheme performs worse with increasing noise. 

\subsubsection{Detector saturation is dominant}
\noindent Here, we want to maximize the compression efficiency [Eq. \eqref{Eq_TotalQFIM}], subject to the constraint that the postselection probability $\PPS$ is no greater than $  P_{\text{max}} \in [0,1]$. We calculate this maximum in Appendix \ref{App_dectector_saturation_calculation}. For notational brevity, we define
\begin{equation}
    b = \left(1-\epsilon+ \dfrac{\epsilon}{d}\right), \quad c = \dfrac{\epsilon}{d}(u-1).
\end{equation}
Mathematically, the optimal filter is most succinctly described by the following five (counting min and max) cases:
\begin{itemize}
    \item{
        $P_{\text{max}} \geq b + c$.\\
        Then the optimal filter has $p_{\bm{\theta}}=B=1$.
    }
    \item {
        $P_{\text{max}} < b + c$. \\
        Letting $t^2=p_{\bm{\theta}}/B$, the optimal filter has
        \begin{align}
            p_{\bm{\theta}} = \frac{P_{\text{max}}}{b + c/t^2} \, , \hspace{1cm}
            B = \frac{P_{\text{max}}}{bt^2 + c}.
        \end{align}
        There are 2 subcases depending on the value of $t_{\text{pp}}$.
        \begin{itemize}
            \item[*] {
            $t_{\text{pp}} \leq 1$. Then, the optimal filter has 
            \begin{equation}
                t^2 = \max\left[t_{\text{pp}}^2, \frac{P_{\text{max}}-c}{b}\right].
            \end{equation}
            }
            \item[*] {
            $t_{\text{pp}} \geq 1$. Then, the optimal filter has 
            \begin{equation}
                t^2 = \min\left[t_{\text{pp}}^2, \left(\frac{c}{P_{\text{max}}-b}\right)^+\right], 
            \end{equation}
            where $a^+ = a$ if $a\geq 0$ and $a^+ = \infty$ if $a<0$.
            }
        \end{itemize}
    }
\end{itemize}

\begin{figure*}
    \centering
    \begin{subfigure}[b]{0.45\textwidth}
        \caption{}\label{Fig_Info_Amp}
        \includegraphics[width = \textwidth]{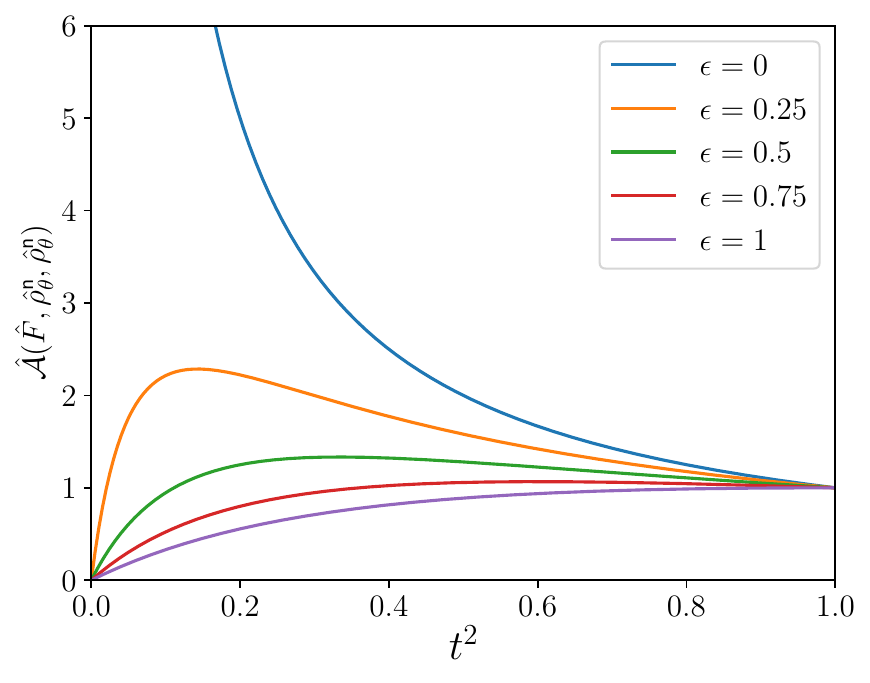}
    \end{subfigure}
    \hfill
    \begin{subfigure}[b]{0.45\textwidth}
        \centering
        \caption{}\label{Fig_Info_Amp_d10_u5}
        \includegraphics[width = \textwidth]{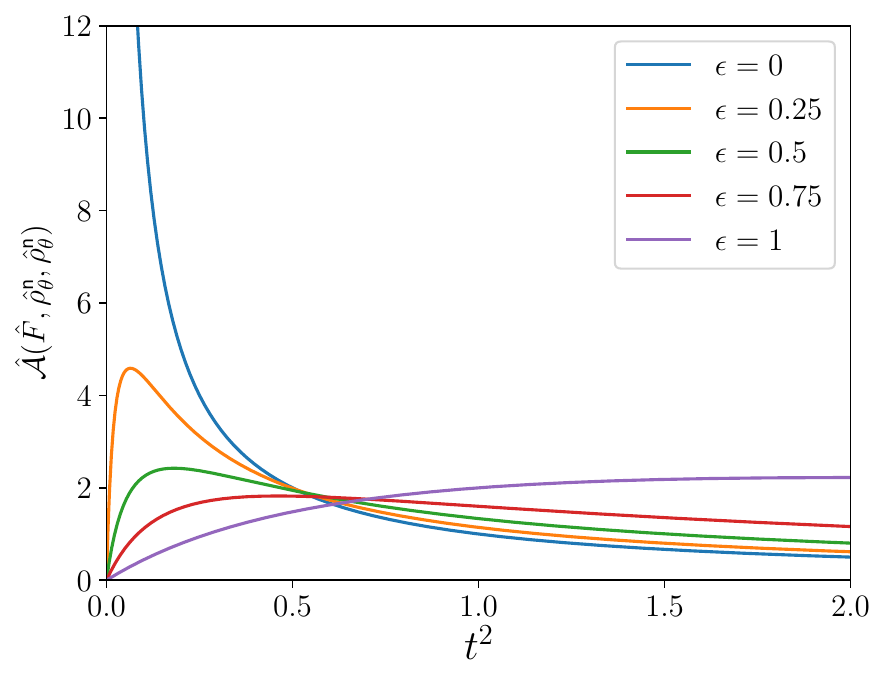}
    \end{subfigure}

    \begin{subfigure}[b]{0.45\textwidth}
        \centering
        \caption{}\label{Fig_max_info_amp_var_d&u}
        \includegraphics[width = \textwidth]{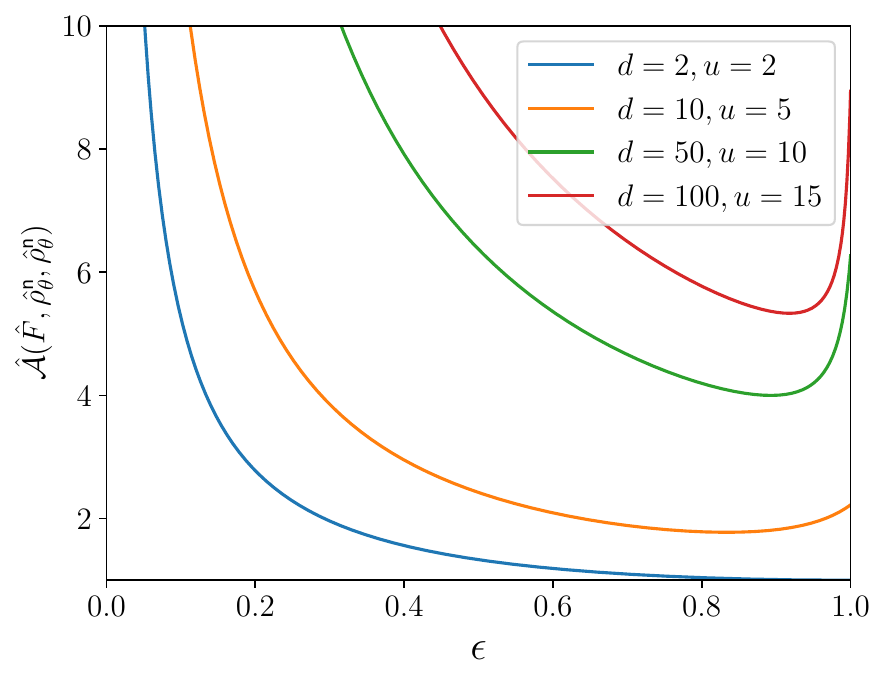}
    \end{subfigure}
    \hfill
    \begin{subfigure}[b]{0.45\textwidth}
        \centering
        \caption{}\label{Fig_info_amp_nonoise}
        \includegraphics[width = \textwidth]{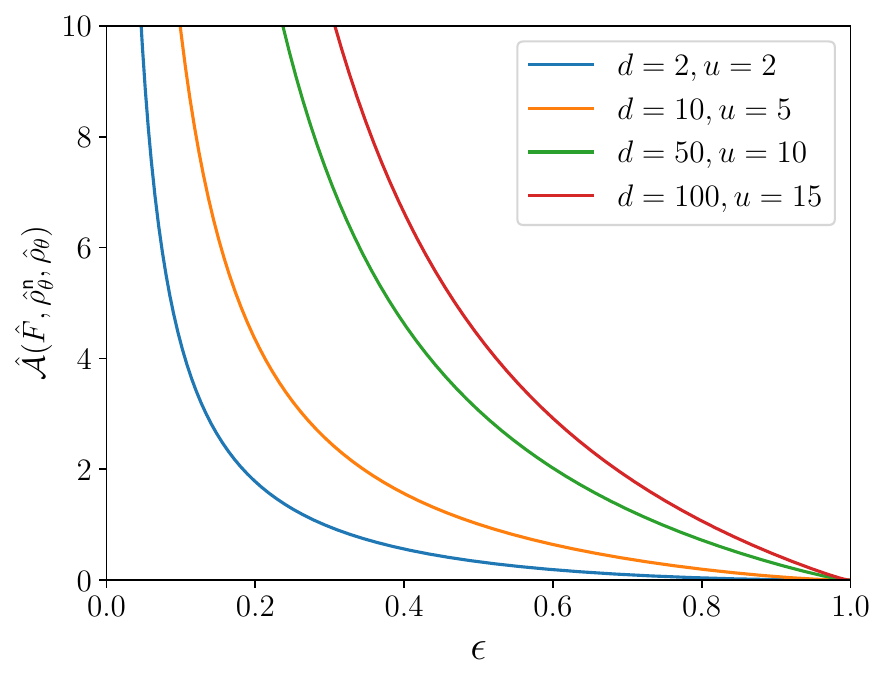}
    \end{subfigure}

    \begin{subfigure}[b]{0.45\textwidth}
        \centering
        \caption{}\label{Fig_comp_efficiency_DS}
        \includegraphics[width = \textwidth]{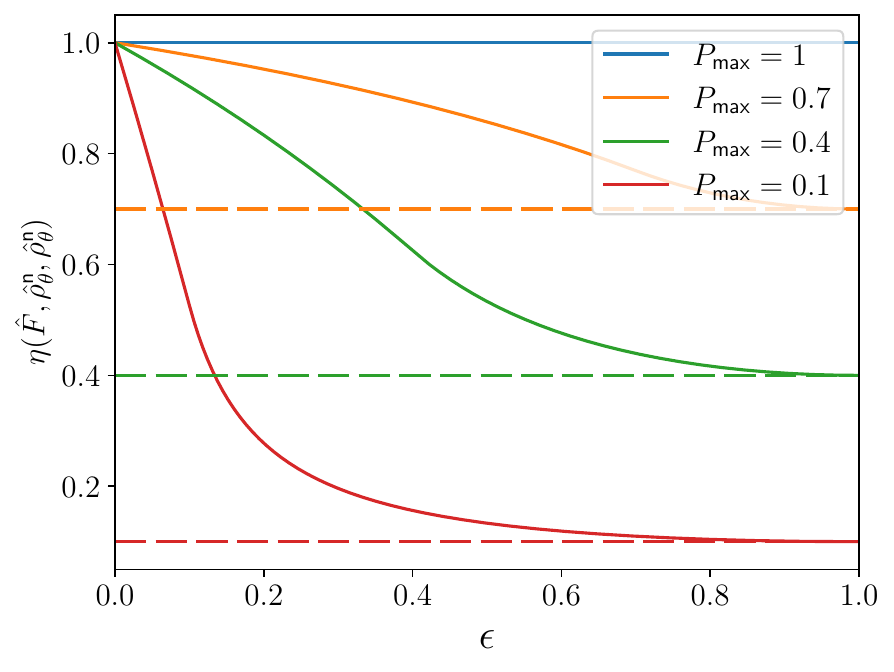}
    \end{subfigure}
    \hfill
    \begin{subfigure}[b]{0.45\textwidth}
        \centering
        \caption{}\label{Fig_comp_efficiency_DS_d10_u5}
        \includegraphics[width = \textwidth]{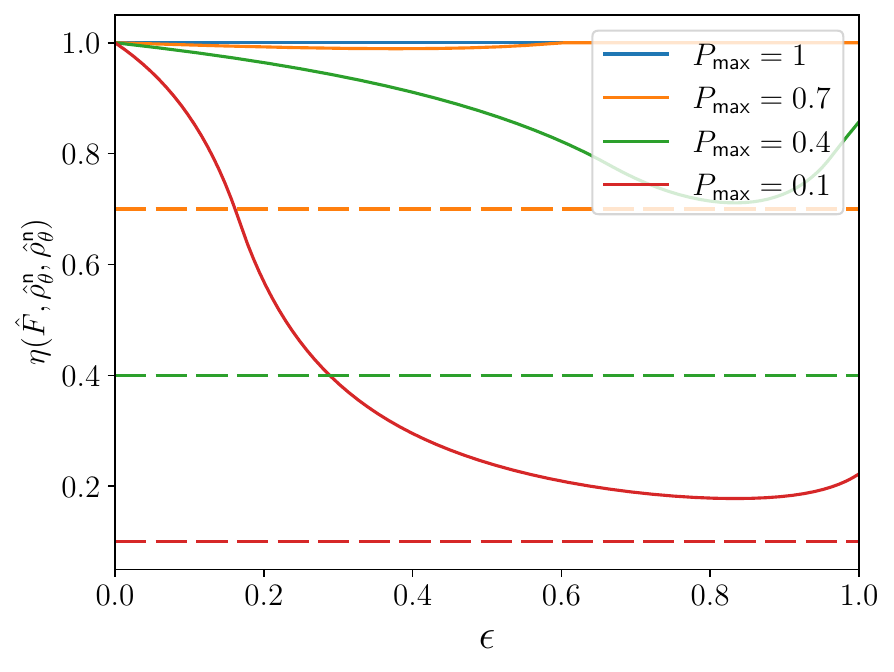}
    \end{subfigure}

    \caption{
        (\subref{Fig_Info_Amp}) Information amplification vs $t^2$ for $d= u = 2$. 
        (\subref{Fig_Info_Amp_d10_u5}) $\hat{\mathcal{A}} ( \state{n,ps}, \state{n})$ vs $t^2$ for $d = 10$ and $u = 5$. 
        %In this case, the optimal $t$ can be larger than $1$ (e.g., for $\epsilon = 1$, $t_{\text{pp}}^2 = 2$). 
        (\subref{Fig_max_info_amp_var_d&u}) Maximum $\hat{\mathcal{A}} (\state{n,ps}, \state{n})$ vs $\epsilon$ for different values of $d$ and $u$. $\hat{\mathcal{A}} ( \state{n,ps}, \state{n})$ reaches a minimum at $\epsilon < 1$, then increases to $d/u$ at $\epsilon = 1$. When compared to an experiment with the same level of noise, our filter performs better for large noise. 
        (\subref{Fig_info_amp_nonoise}) Maximum $\mathcal{A}(\state{n,ps}, \state{})$ vs $\epsilon$ for various values of $d$ and $u$. $\hat{\mathcal{A}} (\state{n,ps}, \state{} )$ is monotonically decreasing with $\epsilon$, reaching $\hat{\mathcal{A}} ( \state{n,ps}, \state{} ) = 0$ at $\epsilon = 1$. As expected, information amplification (with respect to a noiseless experiment) decreases with stronger noise.
        (\subref{Fig_comp_efficiency_DS}) Dashed lines: compression efficiency for the näive classical filter $\hat{F} = P_{\text{max}} \hat{1}$, for $d = 10$, $u = 5$. Solid lines: compression efficiency for the optimal detector saturation filter. Our filter always performs better than the näive classical filter. As $\epsilon \rightarrow 0$ our filter can compress all of the available information into an arbitrarily small number of states.
        (\subref{Fig_comp_efficiency_DS_d10_u5}) Same as Fig. (\subref{Fig_comp_efficiency_DS}), but with $d = 10$ and $u = 5$. Our filter compress all the available information for $P_{\text{max}} > b + c$. For $P_{\text{max}} < b + c$, even as $\epsilon = 1$, our filter performs better than the naive classical filter, by a factor of $d/u$.} 
\label{Fig_amplification}
\end{figure*}

Physically, the optimal filter can fall into one of three distinct categories depending on the size of $P_{\max}$. The three categories arise from grouping the cases above into optimal filters with common behaviors. To describe the categories, we introduce $P_*$ as the largest value of $P_{\text{max}}$ for which the optimal filter has $t^2$ equal to $t_{\text{pp}}^2$. We find that the three different categories are:
\begin{itemize}
    \item {
        $b+c = 1-\epsilon + u\, \epsilon / d  \leq P_{\max}$, whereupon $p_{\bm{\theta}}=B=1$. In this scenario, the filter simply blocks all states that can only arise from noise, and lets through all the states that carry information about the parameters. This regime is lossless (when compared to the information carried in $\hat{\rho}_{\bm{\theta}}^{\text{n}}$).
    }
    \item{
        $P_*\leq P_{\max} < b+c  = 1-\epsilon + u\, \epsilon / d$. In this scenario, the filter compresses information. Since compression with noise is lossy, it does the smallest amount of compression possible, i.e. such that $\PPS = P_{\max}$. It does this by maximizing the information amplification, which corresponds to setting $t$ as close as possible to $t_{\text{pp}}$ whilst still satisfying $\PPS = P_{\max}$. In this regime, the filter incurs some loss due to noise.
    }
    \item{
        $0\leq P_{\max} < P_*$. When $P_{\max}$ reaches $P_*$, it is no longer advantageous to  compress information. Instead, $p_{\bm{\theta}}/B = t_{\text{pp}}^2$ stays constant whilst $\PPS = P_{\max}$, so that both $p_{\bm{\theta}}$ and $B$ decrease. The optimal filter can be decomposed as a compressive filter with $t=t_{\text{pp}}$ followed by a filter that is proportional to $\hat{1}$. In this scenario, the filter compresses some of the information, but is also forced to blindly discard a constant fraction of the states.
    }
\end{itemize}
We can compare our strategy with a näive classical strategy that would blindly discard a constant fraction of the states: 
\begin{equation}
\hat{F}_{\text{nav}} = P_{\max} \, \hat{1} \, . 
\end{equation}
This filter gives a compression efficiency of $P_{\max}$. In Fig. \ref{Fig_comp_efficiency_DS}, we plot the compression efficiency against noise for different values of $P_{\text{max}}$, at $d=u=2$. For small noise ($\epsilon \rightarrow 0$), our postselection strategy can compress information into an arbitrarily small number of measurement probes, while preserving all of the available information. As expected, this advantage decreases and becomes vanishingly small as $\epsilon \rightarrow 1$. In this limit, the performance of our filter reduces to that of the näive classical strategy, which randomly keeps a fraction $P_{\text{max}}$ of the input probes.  In Fig. \ref{Fig_comp_efficiency_DS_d10_u5}, we repeat the same calculation for the case $d=10$ and $u=5$. This time, if $P_{\text{max}} > b + c$, it is possible to preserve all of the available information, even as $\epsilon \rightarrow 1$. For $P_{\text{max}} < b + c$, the situation is similar to the case $d= u = 2$, but our postselection strategy now always performs better than the näive classical strategy.

\section{Conclusion}

\noindent In this paper we derived the family of optimal postselection filters for noiseless multi-parameter quantum metrology.  In the absence of noise, these filters are lossless: they compress information equally across all parameters, while decreasing the postselection probability by the same factor. We showed that the previously-proposed JAL filter \citep{Jenne2021} is contained wthin our family of optimal filters. 

Noise in real experiments leads to natural limits on compression of information. The quantum optics experiment in Ref. \citep{Lupu-Gladstein2021} was affected by errors  in $t^2$, the filter transmission probability, and the unitary operation $\hat{U}(\bm{\theta})$, which bounded the maximum possible compression of information and made the filtering procedure somewhat lossy. This motivated our analysis of the effect of noise on postselected quantum metrology.

We focused our noise-analysis on the worst-case scenario of depolarizing noise of strength $\epsilon$. We considered situations where this noise was applied either before or after the probes have been postselected. When noise acts after the postselection, we found that the JAL filter remains optimal.

We also analyzed the JAL filter's performance when noise acts before the postselection. In this case, we considered two regimes, whereby either post-processing or detector saturation is the main concern. To make the problem tractable, we considered a subset of experimentally sensible filters [Eq. \eqref{eqn:diagonal_F_family}]. To alleviate post-processing costs, one wants to receive the most information in the smallest number of probes, which corresponds to maximizing the information amplification [Eq. \eqref{Eq_QFIMamplification}]. To protect against detector saturation, we fixed a maximum postselection probability $P_{\text{max}}$ and maximized the compression efficiency [Eq. \eqref{Eq_TotalQFIM}]. We showed that the JAL filter is not always optimal, but with slight modifications it performs well in both the post-processing and detector saturation regimes. Thus, if one can estimate the strength of noise in an experiment, one should implement a different filter compared to the case of no noise.

We showed that for qubit sensors ($u=d=2$) the family of filters in Eq. \eqref{Eq_filters_considered} is optimal. However, for finite $u,d \neq 2$, we show that this subset of filters is not always optimal. In fact, it can sometimes be advantageous for states corresponding to noise to be mixed with states corresponding to changes in $\bm{\theta}$. 

Wit this work, we hope to extend the use cases of postselected metrology to non-optical systems.

\section*{Acknowledgments}
\noindent The authors wish to thank N.Y. Halpern, N. Gladstein and Y. Batuhan Yilmaz for their useful discussions and comments. F.S. was supported by the Harding Foundation and the Cambridge Vice-Chancellor's Award, W.S. was supported by the EPSRC and Hitachi, D.R.M.A.-S. was supported by the Lars Hierta’s Memorial Foundation and Girton College.

\bibliographystyle{apsrev4-1}
\bibliography{apssamp.bib}

\clearpage
\appendix

\onecolumngrid

\section{Proof of optimal noiseless filter} \label{Appendix_Theorem}
In this Appendix we prove the main Theorem from section \ref{Sec_OPTFilt}. \\

From Eq. \eqref{Eq_Noiseless_QFIM}, we see that the QFIM only contains terms that depend on $\hat{F}_u$, and thus we deduce part 1 of the theorem. 

\par We now focus on the diagonal entries of the QFIM: $\mathcal{I}_{j,j}({\bm{\theta}} | \hat{\rho}_{\bm{\theta}} ^{\text{ps}})$. Taking the derivative of $\innerproduct{\psi_{\bm{\theta}}}{\psi_{\bm{\theta}}} \equiv 1$, we see that:
\begin{equation}
    \textrm{Re}(\innerproduct{\partial_j \psi_{\bm{\theta}}}{\psi_{\bm{\theta}}} ) = 0 \, .
\end{equation}
Thus we can write
\begin{equation} \label{Eq_PerpStates}
    \ket{ \partial_j\psi_{\bm{\theta}}} = i x_j \ket{\psi_{\bm{\theta}}} + \alpha_j \ket{{\psi_{\bm{\theta}}}^{\perp,j}} \, ,
\end{equation}
for $x_j \in\mathbb{R}$ a \textit{real} coefficient, $\alpha_j \in \mathbb{C}$ and some normalized set of $\ket{\psi^{\perp,j}}$, orthogonal to $\ket{\psi_{\bm{\theta}}}$ (note that they may not be orthogonal to each other). Define:
\begin{align}
    \expval{\hat{F}}{\psi_{\bm{\theta}}} &= \PPS \in [0,1] \, ,\\
    \braket{{\psi_{\bm{\theta}}}^{\perp,j}| \hat{F} | {\psi_{\bm{\theta}}}^{\perp,j}} &= B_j \in [0,1] \, ,\\
    \braket{\psi_{\bm{\theta}} | \hat{F} | {\psi_{\bm{\theta}}}^{\perp,j}} &= C_j \in\mathbb{C} \, .
\end{align}
Substituting these expressions into Eq. \eqref{Eq_Noiseless_QFIM}, we find: 
\begin{align}
    \mathcal{I}_{j,j}({\bm{\theta}} | \hat{\rho}_{\bm{\theta}} ^{\text{ps}}) & = 4\Bigg[ \dfrac{1}{\PPS} \Big(  x_j ^2 \PPS + |\alpha_j|^2 B_j + i x_j (\alpha_j^*C_j ^*-\alpha_jC_j)  \Big)  - \dfrac{1}{({\PPS})^2} |  i x_j \PPS  + \alpha_j C_j |^2 \Bigg] \, , \\
    &= 4 \Big( \dfrac{|\alpha_j|^2 B_j}{\PPS} - \dfrac{|\alpha_j C_j|^2}{({\PPS})^2} \Big) \, .
\end{align}
Setting $\hat{F}=\hat{1}$, we have:
\begin{equation}
    \mathcal{I}_{j,j}({\bm{\theta}} | \hat{\rho}_{\bm{\theta}}) = 4 |\alpha_j|^2.
\end{equation}
It therefore follows that:
\begin{equation} \label{Eq_Noiseless_QFIM_amplification}
     \mathcal{A}(\state{ps}, \state{}) =  \Big( \dfrac{B_j}{\PPS} - \dfrac{|C_j|^2}{\, ({\PPS})^2} \Big) \, .
\end{equation}
In order for $ \mathcal{A}(\state{ps}, \state{}) = 1/\PPS$, we require that $C_j = 0$ (by assumption $\mathcal{I}_{j,j}({\bm{\theta}} | \hat{\rho}_{\bm{\theta}}) \, , \alpha_j  \neq 0$) and $B_j = 1$. We must therefore have:
\begin{equation}
    \hat{F}_u = (\PPS-1)\dyad{\psi_{\bm{\theta}}} + \hat{\Pi}_u \, .
\end{equation} 
With this, we have deduced uniqueness of an optimal $\hat{F}_u$. In Ref. \citep{Jenne2021} it was shown that this $\hat{F}_u$ is indeed lossless (which requires calculating the off diagonal terms). $\square$

It may be that $\mathcal{U}$ has a dimensionality less than $d$, the dimensionality of the unitary $U(\bm{\theta})$. For example, this will always happen if $M+1<d$, where $M$ is the number of parameters. In general, $\mathcal{U}$ can have dimension $u$, with $u \leq M+1$, and if $u < d$, the optimal filter is not fully determined. In an orthonormal basis, where $\ket{\psi_{\bm{\theta}}}$ is the first basis vector, and the first $u$ basis vectors span $\mathcal{U}$, we have that the optimal filter takes the form
\begin{equation} \label{Eq_General_Noiseless_Filter}
    \hat{F} = \begin{pmatrix}
        \begin{pmatrix}
            \PPS & \dots & 0 \\
             \vdots & \ddots & \vdots \\
             0 & \dots & 1
        \end{pmatrix}
         & \hat{C} \\
         \hat{C}^{\dagger} & \hat{D}
    \end{pmatrix} \, ,
\end{equation}
where $\hat{C}$ and $\hat{D}$ are matrices such that $\hat{0} \leq \hat{F} \leq \hat{1}$.

\par The simplest example of an optimal postselection filter is for $\hat{C} = 0$ and $ \hat{D} =\hat{1}$, in which case, setting $\PPS = t^2$, we recover the JAL filter: 
\begin{equation}
    \hat{F} = (t^2-1) \hat{\rho}_{\bm{\theta}} + \hat{1} \, .
\end{equation}
Note that the JAL filter does not require knowledge of $\mathcal{U}$. 

So far, we have assumed that our filter is constructed based on perfect knowledge of the parameters, i.e. our initial estimate $\bm{\theta}_0$ is equal to the true value of the parameters $\bm{\theta}$. In practice, however, one does not know $\bm{\theta}$ exactly. Instead, one has an estimate $\bm{\theta}_0 = \bm{\theta-\delta}$ and implements a filter using $\bm{\theta}_0$. In Ref. \cite{Jenne2021}, it was shown that if one implements the JAL filter using $\bm{\theta}_0$, then $\mathcal{A}(\state{ps}, \state{})$ and $\PPS$ are only perturbed at order $\bm{\delta}^2$ - there is no order $\bm{\delta}$ correction.

In Appendix \ref{App_JAL_Noiseless_Scaling}, we study the order $\bm{\delta}^2$ correction when using the JAL filter, deriving the upper bound
\begin{equation} \label{Eq_AppD_PSInfo}
     \mathcal{A}(\state{ps}, \state{})  \leq  \dfrac{1}{t^2}  \Big[ 1- \dfrac{1-t^2}{t^2} \,  \bm{\delta}^T  \mathcal{I}(\bm{\theta} | \hat{\rho}_{\bm{\theta}}) \, \bm{\delta}  \,   \Big]   \, .
\end{equation}
Equation \eqref{Eq_AppD_PSInfo} provides a way to calculate the expected decrease in the information amplification to second order in $\bm{\delta}$. It allows one to understand which parameters in $\bm{\delta}$ are responsible for the largest loss of information. In the trivial case when $\mathcal{I}(\bm{\theta} | \hat{\rho}_{\bm{\theta}})$ is diagonal, the parameters $\theta_i$ with the largest QFI will also lead to the greatest loss.

\section{Information amplification scaling for the JAL filter} \label{App_JAL_Noiseless_Scaling}
\noindent In this Appendix, we study the second order corrections to the information amplification when $\bm{\delta}\neq \bm{0}$. Consider the perturbed JAL filter
\begin{equation}
\hat{F} = (t^2 -1 ) \hat{\rho}_{\bm{\theta}_0} + \hat{1} \, .
\end{equation}
We expand $\hat{U}(\bm{\theta}_0)$ to second order in $\bm{\delta}$:
\begin{equation}
        \hat{U}(\bm{\theta}_0) = \hat{U}(\bm{\theta}) + \Big[ \partial_{i} \hat{U}(\bm{\theta}) \Big] (\bm{\theta}_0 - \bm{\theta})_i + \dfrac{1}{2} \Big[\partial^2_{i,j} \hat{U}(\bm{\theta}) \Big] (\bm{\theta}_0 - \bm{\theta})_i (\bm{\theta}_0 - \bm{\theta})_j \, .
\end{equation}
The above equation can be written as
\begin{align}
  \hat{U}(\bm{\theta}_0) = U(\bm{\theta}) + \Big[ \nabla_{\bm{\theta}} \hat{U}(\bm{\theta}) \Big]^{T} (\bm{\theta}_0 - \bm{\theta}) + \dfrac{1}{2} (\bm{\theta}_0 - \bm{\theta})^{T} H(\bm{\theta})  \,  (\bm{\theta}_0 - \bm{\theta})  \, ,
\end{align}
where $\hat{H}(\theta) = \partial^2 _{i,j} U(\bm{\theta})$ is the Hessian matrix. Hence
\begin{align}
  \hat{U}(\bm{\theta}_0) = \hat{U}(\bm{\theta}) - i \hat{U}(\bm{\theta}) \hat{d}_1 + \dfrac{i}{2}  \hat{U}(\bm{\theta}) \hat{d}_2  + \mathcal{O}(\bm{\delta}^3)\, ,
\end{align}
where $\hat{d}_1$ and $\hat{d}_2$ are defined as
\begin{align}
    \hat{d}_1 & = -i \, \hat{U}^{\dagger}(\bm{\theta}) \Big[  \nabla_{\bm{\theta}} \hat{U}(\bm{\theta}) \Big]^{T} \bm{\delta} \, , \\
    \hat{d}_2 & = - i \, \hat{U}^{\dagger}(\bm{\theta}) \, \bm{\delta}^{T} \hat{H}(\bm{\theta}) \, \bm{\delta} \, .
\end{align}
We want to find $\hat{d}^{\dagger}_1$ and $\hat{d}^{\dagger}_2$. Consider
\begin{equation}
    U(\bm{\theta}_0) U(\bm{\theta}_0)^{\dagger} = \hat{1} \, .
\end{equation}
Inserting the Taylor expansion, we find
\begin{align}
      \hat{1} & = \Big[ \hat{U}(\bm{\theta}) - i \hat{U}(\bm{\theta}) \hat{d}_1 + \dfrac{i}{2}  \hat{U}(\bm{\theta}) \hat{d}_2 \Big] \Big[\hat{U}(\bm{\theta}) - i \hat{U}(\bm{\theta}) \hat{d}_1 + \dfrac{i}{2}  \hat{U}(\bm{\theta}) \hat{d}_2\Big]^{\dagger} \\
      & = \hat{U}(\bm{\theta})  \hat{U}^{\dagger}(\bm{\theta}) + i \hat{U} (\bm{\theta}) \Big[  \hat{d}_1  - \hat{d}_1^{\dagger} \Big] \hat{U}^{\dagger} (\bm{\theta}) + \dfrac{i}{2} \hat{U} (\bm{\theta}) \Big[  \hat{d}_2 -   \hat{d}_2^{\dagger} - 2 \, i \,  \hat{d}_1 \hat{d}^{\dagger}_1   \Big] \hat{U}^{\dagger} (\bm{\theta}) \, .
\end{align}
Looking at the term of order $\bm{\delta}$, we immediately see that $\hat{d}_1 = \hat{d}^{\dagger}_1$. Instead, the second order term gives
\begin{equation}
    \hat{d}_2^{\dagger} =  \hat{d}_2 - 2 \, i \, \hat{d}_1 \hat{d}^{\dagger}_1  \, .
\end{equation}
Now, we can write $\hat{\rho}_{\bm{\theta}_0}$ as
\begin{align}
    \hat{\rho}_{\bm{\theta}_0} & = \hat{U}(\bm{\theta}_0) \rho_0 \hat{U}(\bm{\theta}_0)^{\dagger} \\
    & = \Big[ \hat{U}(\bm{\theta}) - i \hat{U}(\bm{\theta}) \hat{d}_1  + \dfrac{i}{2} \hat{U}(\bm{\theta}) \hat{d}_2  \Big]  \rho_0 \Big[\hat{U}(\bm{\theta}) ^{\dagger }+ i  \hat{d}^{\dagger}_1  \hat{U}(\bm{\theta})^{\dagger} - \dfrac{i}{2} \hat{d}^{\dagger}_2 \hat{U}(\bm{\theta})^{\dagger} \Big] \\
    & =  \hat{U}(\bm{\theta}) \Big[ 1 - i \hat{d}_1  + \dfrac{i}{2}  \hat{d}_2  \Big]  \rho_0 \Big[1+ i  \hat{d}_1   - \dfrac{i}{2} \hat{d}_2 - \hat{d}^2_1\Big] \hat{U}(\bm{\theta}) ^{\dagger }\, , \\
    & = \hat{\rho}_{\bm{\theta}} + i [\hat{\rho}_{\bm{\theta}}, \hat{D}_1] - \dfrac{i}{2} [ \hat{\rho}_{\bm{\theta}}, \hat{D}_2 ]  + \hat{D}_1 \hat{\rho}_{\bm{\theta}} \hat{D}_1 -  \hat{\rho}_{\bm{\theta}} \hat{D}^2 _1 +  \mathcal{O}(\bm{\delta}^3)  \, ,
\end{align}
where we defined $\hat{D}_1, \hat{D}_2$:
\begin{align}
\hat{D}_1 = U(\bm{\theta}) \hat{d}_1 \hat{U}(\bm{\theta})^{\dagger}  \, , \\
\hat{D}_2 = U(\bm{\theta}) \hat{d}_2 \hat{U}(\bm{\theta})^{\dagger}  \, . 
\end{align}
We can now calculate the postselection probability $\PPS$:
\begin{align}
    \PPS & = \text{Tr}[ \hat{F} \hat{\rho}_{\bm{\theta}}] \, ,  \\
    &= \braket{\psi_{\bm{\theta}} | \hat{F} | \psi_{\bm{\theta}}} \, ,  \\
    &= \braket{\psi_{\bm{\theta}} | \Big[ (t^2 -1) \Big( \hat{\rho}_{\bm{\theta}} + i [\hat{\rho}_{\bm{\theta}}, \hat{D}_1] -  \dfrac{i}{2} [ \hat{\rho}_{\bm{\theta}}, \hat{D}_2 ]  + \hat{D}_1 \hat{\rho}_{\bm{\theta}} \hat{D}_1 -  \hat{\rho}_{\bm{\theta}} \hat{D}^2 _1 +  \mathcal{O}(\bm{\delta}^3) \Big) + \hat{1} \Big] | \psi_{\bm{\theta}}}  + \mathcal{O}(\bm{\delta}^3) \, ,  \\
    &= t^2 -  (1-t^2) \Big[ \braket{\psi_{\bm{\theta}} |  \big( \hat{D}_1 \hat{\rho}_{\bm{\theta}} \hat{D}_1 \big) | \psi_{\bm{\theta}}} -  \braket{ \psi_{\bm{\theta}} | \hat{\rho}_{\bm{\theta}} \hat{D}^2 _1 | \psi_{\bm{\theta}}} \Big] + \mathcal{O}(\bm{\delta}^3)  \, ,  \\
    &= t^2 +  (1- t^2) \Big[  \braket{\psi_{\bm{\theta}} | \hat{D}^2 _1 | \psi_{\bm{\theta}}}    - | \braket{\psi_{\bm{\theta}} | \hat{D}_1  | \psi_{\bm{\theta}}} | ^2 \Big]   + \mathcal{O}(\bm{\delta}^3) \, ,  \\
    & = t^2 +  (1- t^2) \Big[  \braket{\psi_{0} | \hat{d}^2_1 | \psi_{0}}   - | \braket{\psi_{0} | \hat{d}_1 | \psi_{0}} | ^2 \Big]  + \mathcal{O}(\bm{\delta}^3) \, , \\
    & = t^2 +  (1- t^2) \sum_{i,j} {\bm{\delta}_i \bm{\delta}_j \Big[  \braket{\partial_i \psi_{\bm{\theta}} | \partial_j \psi_{\bm{\theta}}} - \braket{\partial_i \psi_{\bm{\theta}} | \psi_{\bm{\theta}}} \braket{ \psi_{\bm{\theta}} | \partial_j \psi_{\bm{\theta}}} } \Big] + \mathcal{O}(\bm{\delta}^3) \, , \\
    & = t^2 +  (1- t^2) \sum_{i,j} {\bm{\delta}_i \bm{\delta}_j \, \mathcal{I} (\bm{\theta} |     \hat{\rho}_{\bm{\theta}})_{i,j}}  + \mathcal{O}(\bm{\delta}^3) \, , \\
    & = t^2 +  (1- t^2) \, \bm{\delta}^{T} \, \mathcal{I} (\bm{\theta} |     \hat{\rho}_{\bm{\theta}}) \, \bm{\delta}  + \mathcal{O}(\bm{\delta}^3) \, .
\end{align}
Hence, we find:
\begin{equation} \label{Eq_PS_Prob_delta}
     \PPS = t^2 + (1-t^2) \, \bm{\delta}^{T} \,  \mathcal{I}(\bm{\theta} | \hat{\rho}_{\bm{\theta}}) \, \bm{\delta}  + 
    \mathcal{O}(\bm{\delta}^3)  \, .
\end{equation}
Since the QFIM is positive semi-definite, the term of order $\mathcal{O}(|\bm{\delta}|^2)$ is always positive. Therefore, for $\bm{\delta}$ small, the postselection probability increases with increasing $|\bm{\delta}|$.

\par We can now find an upper bound on $\mathcal{A}(\state{ps}, \state{})$. From the discussion in Ref. \cite{Jenne2021}, we know that 
\begin{equation} \label{Eq_QFIM_delta}
     \mathcal{I} (\bm{\theta} |     \hat{\rho}_{\bm{\theta}} ^{ \text{ps}})  =  \dfrac{1}{t^2}  \Big[ \mathcal{I} (\bm{\theta} |     \hat{\rho}_{\bm{\theta}})  + \Delta \Big] \, ,
\end{equation}
where $\Delta$ is a matrix of order $\mathcal{O}(|\bm{\delta}|^2)$ which we wish to bound. Because  $\eta(\PPS,\state{ps}, \state{}) \leq 1$, we can write
\begin{equation} \label{Eq_QFIM_Ineq}
    \PPS \, \mathcal{I} (\bm{\theta} |     \hat{\rho}_{\bm{\theta}} ^{ \text{ps}}) \leq \mathcal{I} (\bm{\theta} |     \hat{\rho}_{\bm{\theta}}) \, .
\end{equation}
Substituting Eqs. \eqref{Eq_PS_Prob_delta} and Eq. \eqref{Eq_QFIM_delta} into Eq. \eqref{Eq_QFIM_Ineq}, we find
\begin{gather*}
\Big[ t^2 + (1-t^2) \,  \bm{\delta}^T  \mathcal{I}(\bm{\theta} | \hat{\rho}_{\bm{\theta}}) \, \bm{\delta} + \mathcal{O}(\bm{\delta}^3) \Big] \dfrac{1}{t^2} \Big[  \mathcal{I} (\bm{\theta} |     \hat{\rho}_{\bm{\theta}})  +  \Delta   \Big] \leq \mathcal{I} (\bm{\theta} |     \hat{\rho}_{\bm{\theta}}) \\
     \mathcal{I} (\bm{\theta} |     \hat{\rho}_{\bm{\theta}}) + \Delta + \dfrac{1-t^2}{t^2}   \bm{\delta}^T  \mathcal{I}(\bm{\theta} | \hat{\rho}_{\bm{\theta}}) \, \bm{\delta} \, \mathcal{I}(\bm{\theta} | \hat{\rho}_{\bm{\theta}}) \leq  \mathcal{I} (\bm{\theta} |     \hat{\rho}_{\bm{\theta}})  + \mathcal{O}(|\bm{\delta}|^3) \\
     \Rightarrow \Delta \leq - \dfrac{1-t^2}{t^2} \big[ \bm{\delta}^T  \mathcal{I}(\bm{\theta} | \hat{\rho}_{\bm{\theta}}) \, \bm{\delta}  \big] \, \mathcal{I} (\bm{\theta} |     \hat{\rho}_{\bm{\theta}}) \,  + \mathcal{O}(|\bm{\delta}|^3) \, .
\end{gather*}
Therefore, we reach the important result
\begin{equation} 
     \mathcal{I} (\bm{\theta} |     \hat{\rho}_{\bm{\theta}} ^{ \text{ps}})  \leq  \dfrac{1}{t^2}  \mathcal{I} (\bm{\theta} |     \hat{\rho}_{\bm{\theta}}) \Big[ 1- \dfrac{1-t^2}{t^2} \,  \bm{\delta}^T  \mathcal{I}(\bm{\theta} | \hat{\rho}_{\bm{\theta}}) \, \bm{\delta}  \,   \Big]  \, ,
\end{equation}
which shows that any $\bm{\delta}\neq 0$ decreases the information amplification.

\section{Noisy QFIM calculations}
\noindent In this Appendix, we calculate the QFIM when noise acts before or after filtering. To calculate the QFIM, we use the explicit expression from Ref. \cite{Liu2019}:
\begin{equation} \label{Eq_QFIM}
   \mathcal{I} (\bm{\theta}|\hat{\rho}_{\bm{\theta}})_{i,j} = 2 \sum_{\substack{n,m =1 \\ \lambda_n + \lambda_m > 0}}^{d}{\dfrac{ \braket{\lambda_n | \partial_{j} \hat{\rho}_{\bm{\theta}} | \lambda_m} \braket{\lambda_m | \partial_{i} \hat{\rho}_{\bm{\theta}} | \lambda_n}     }{\lambda_n + \lambda_m} }  \, .
\end{equation}.

\subsection{Noise after postselection} \label{Appendix_NoiseAfter}
\noindent We start by calculating the QFIM of $\hat{\rho}^{\text{n}}_{\bm{\theta}} $ using Eq. \eqref{Eq_QFIM}, ignoring postselection for the moment. We calculate the eigenvalues $\{\lambda_i\}$ and eigenvectors $\{\ket{\lambda_i}\}$ of $\hat{\rho}^{\text{n}}_{\bm{\theta}}$:
\begin{align}
    \lambda_1 = (1-\epsilon) + \dfrac{\epsilon}{d} & \, , \quad  \lambda_{i} = \dfrac{\epsilon}{d} \, \,  \text{for } i\neq1 \, , \\
   \ket{\lambda_1} =  \ket{\psi_{\bm{\theta}}} &\, , \quad \ket{\lambda_{i}}  =  \ket{\psi_{\bm{\theta}}^{\perp, i}} \, \text{for } i\neq1\, , 
\end{align}
where we have defined an orthonormal eigenbasis $\{ \ket{\psi_{\bm{\theta}}}, \ket{{\psi_{\bm{\theta}}}^{\perp, 2}}, \dots, \ket{{\psi_{\bm{\theta}}}^{\perp, d}}\}$ and $1 \leq i \leq d $.

\par Substituting $\partial_{i} \hat{\rho}^{\text{n}}_{\bm{\theta}}$ in Eq. \eqref{Eq_QFIM}, where
\begin{equation}
    \partial_{i} \hat{\rho}^{\text{n}}_{\bm{\theta}} = (1-\epsilon) \big( \ket{\partial_{i}{\psi_{\bm{\theta}}}} \bra{\psi_{\bm{\theta}}} + \ket{\psi_{\bm{\theta}}} \bra{\partial_{i}{\psi_{\bm{\theta}}}} \big) \, ,
\end{equation}
we see that the term with $n = m = 1$ is zero, since $\braket{\partial_{i}{\psi_{\bm{\theta}}} | \psi_{\bm{\theta}}}  + \braket{\psi_{\bm{\theta}} |\partial_{i}{\psi_{\bm{\theta}}}} = \partial_{i}(\braket{\psi_{\bm{\theta}} | \psi_{\bm{\theta}}}) = \partial_{i}{(1)} = 0$. Similarly, the terms with both $n, m \neq 1$ are also zero, since $\braket{\psi_{\bm{\theta}} | {\psi_{\bm{\theta}}}^{\perp, n}} =0$. The only non-zero terms are for $n \neq 1$ and $m = 1$ (or vice-versa), giving:
\begin{align}
    \hspace{0cm} \mathcal{I}_{i,j}(\bm{\theta} | \hat{\rho}^{\text{n}}_{\bm{\theta}}) &= 4 (1-\epsilon)^2 \sum_{n \neq 1}{\dfrac{ \text{Re}\Big[ \braket{\partial_i{\psi_{\bm{\theta}}} | \lambda_n} \braket{\lambda_n|\partial_j{\psi_{\bm{\theta}}}}   \Big] }{\lambda_n + \lambda_1}} \, , \\
      & \hspace{-1cm} = 4 (1-\epsilon)^2 {\dfrac{ \text{Re}\Big[ \bra{\partial_i{\psi_{\bm{\theta}}}} (\hat{1} - \ket{\lambda_1} \bra{\lambda_1})\ket{\partial_j{\psi_{\bm{\theta}}} } \Big] }{\lambda_{2} + \lambda_1}} \, ,\\
      & \hspace{-1cm} = {\dfrac{4 (1-\epsilon)^2 }{(1-\epsilon) +   \dfrac{2\epsilon}{d}}} \text{Re} \Big[ \braket{\partial_i{\psi_{\bm{\theta}}}| \partial_j{\psi_{\bm{\theta}}}} - \braket{\partial_i{\psi_{\bm{\theta}}} |\psi_{\bm{\theta}}}  \braket{\psi_{\bm{\theta}} | \partial_j{\psi_{\bm{\theta}}}}  \Big] \, , \\
    & \hspace{-1cm} = {\dfrac{(1-\epsilon)^2 }{(1-\epsilon) +   \dfrac{2\epsilon}{d}}} \, \mathcal{I}_{i,j}(\bm{\theta} | \hat{\rho}_{\bm{\theta}})   \label{Eq_finalline}   \, .
\end{align}
Since we have not assumed any particular form for the pure state $\ket{\psi_{\bm{\theta}}}$, we can postselect it with a general 2-outcome POVM $\{ \hat{F}_1 = \hat{F} , \hat{F}_2 = 1- \hat{F}  \}$. Blocking the states corresponding to outcome $\hat{F}_2$ and letting through those corresponding to outcome $\hat{F}_1 = \hat{F}$, we find that the postselected state is
\begin{equation}
    \ket{{\psi_{\bm{\theta}}}^{\text{ps}}} = \dfrac{\hat{K} \ket{\psi_{\bm{\theta}}}}{\sqrt{\PPS}}, \quad \text{where} \quad \PPS = \text{Tr} \, \Big[\hat{F} \dyad{\psi_{\bm{\theta}}} \Big] \, ,
\end{equation}
where $\hat{F} = \hat{K}^{\dagger} \hat{K}$. Hence,
\begin{equation} \label{Eq_QFIM_Noise_Before}
\mathcal{I}(\bm{\theta} | \hat{\rho}_{\bm{\theta}} ^{\text{ps, n}} ) = {\dfrac{ (1-\epsilon)^2 }{(1-\epsilon) +  \dfrac{2 \epsilon}{d}}}  \mathcal{I}(\bm{\theta} | \hat{\rho}_{\bm{\theta}} ^{\text{ps}})\, .
\end{equation}
At the same time, from the result in Eq. \eqref{Eq_finalline}:
\begin{equation} \label{Eq_finalline_repeat}
\mathcal{I}(\bm{\theta} | \hat{\rho}^{\text{n}}_{\bm{\theta}}) = {\dfrac{ (1-\epsilon)^2 }{(1-\epsilon) +  \dfrac{2 \epsilon}{d}}}  \mathcal{I}(\bm{\theta} | \hat{\rho}_{\bm{\theta}})  \, .
\end{equation}
Taking the ratio of Eqs. \eqref{Eq_QFIM_Noise_Before} and \eqref{Eq_finalline_repeat}, and using $\mathcal{I}(\bm{\theta} | \hat{\rho}_{\bm{\theta}}^{\text{ps}} )  \approx \mathcal{I}(\bm{\theta} | \hat{\rho}_{\bm{\theta}})/ t^2 + \mathcal{O}(|\bm{\delta}|^2 )$, we find that
\begin{equation}
    \mathcal{I}(\bm{\theta} | \hat{\rho}_{\bm{\theta}}^{\text{ps, n}} ) = \dfrac{1}{t^2} \mathcal{I}(\bm{\theta} | \hat{\rho} ^{\text{n}} _{\bm{\theta}}  ) + \mathcal{O}(|\bm{\delta}|^2 ) \, .
\end{equation}

Hence all of the loss of information comes from noise, and filtering is still lossless.

\subsection{Noise before postselection} \label{Appendix_NoiseBefore}
\noindent When noise acts before postselection, the resulting noisy postselected state is
\begin{equation} \label{Eq_rho}
        \hat{\rho}_{\bm{\theta}} ^{\text{n, ps}} =  \dfrac{1}{\PPS} \Big[ (1-\epsilon) \hat{K} \hat{\rho}_{\bm{\theta}} \hat{K}^\dag +  \dfrac{\epsilon}{d} \,  \hat{K}\hat{K}^\dag \Big] \, ,
\end{equation}
where the normalization constant $\PPS$ is given by
\begin{equation}
     \PPS = (1-\epsilon) \, \text{Tr} \Big[ \hat{F} \hat{\rho}_{\bm{\theta}} \Big] + \dfrac{\epsilon}{d}  \, \text{Tr} \Big[ \hat{F} \Big] \, .
\end{equation}
We consider filters of the form
\begin{equation}
    \hat{F} = (p_{\bm{\theta}}-B)\dyad{\psi_{\bm{\theta}}} + B \, \hat{\Pi}_u + D\, \hat{\Pi}_n\, ,
\end{equation}
where $p_{\bm{\theta}},B,D\in[0,1]$, $\hat{\Pi}_u$ is the orthogonal projection onto $\mathcal{U}$ and $\hat{\Pi}_n$ is the orthogonal projection onto $\mathcal{U}^\perp$. For suitable bases, this filter is represented by the matrix:

\begin{equation}\label{eqn:diagonal_F_family}
    \hat{F} = \begin{pmatrix}
        \begin{pmatrix} 
            p_{\bm{\theta}} & \dots & 0 \\
             \vdots & \ddots & \vdots \\
             0 & \dots & B
        \end{pmatrix}
         & 0 \\
         0 & \begin{pmatrix}
            D & \dots & 0 \\
             \vdots & \ddots & \vdots \\
             0 & \dots & D
        \end{pmatrix}
    \end{pmatrix} \, .
\end{equation}

\noindent We  start by making the assumption that $\bm{\delta} =0$, i.e. $\bm{\theta}_0 = \bm{\theta}$. In Appendix \ref{Appendix_C}, we then show that the result is unchanged for small $\bm{\delta}$, up to a term of order $\mathcal{O}(|\bm{\delta}|^2 )$. 
 
\par Before calculating the QFIM using Eq. \eqref{Eq_QFIM}, we need to find the eigenvalues and eigenvectors of $\hat{\rho}_{\bm{\theta}} ^{\text{n, ps}}$. We first look at the eigenvalues and eigenvectors of $\hat{\rho}_{\bm{\theta}} ^{\text{n, ps}}$. Letting $u=\dim \, \mathcal{U}$, we can pick orthonormal bases
$\{ \ket{\psi_{\bm{\theta}}}, \ket{{\psi_{\bm{\theta}}}^{\perp, 2}}, \dots, \ket{{\psi_{\bm{\theta}}}^{\perp, u}}\}$ of $\mathcal{U}$, and  $\{\ket{{\psi_{\bm{\theta}}}^{\perp, u+1}}, \dots, \ket{{\psi_{\bm{\theta}}}^{\perp, d}}\}$ of $\mathcal{U}^\perp$ that satisfy:
\begin{align}
    \hat{F} \ket{\psi_{\bm{\theta}}} &= p_{\bm{\theta}}\ket{\psi_{\bm{\theta}}},\\
    \quad\hat{F}\ket{{\psi_{\bm{\theta}}}^{\perp, i}} &=  B \ket{{\psi_{\bm{\theta}}}^{\perp, i}} \text{ for }  2 \leq i \leq u \, , \\
    \quad \hat{F}\ket{{\psi_{\bm{\theta}}}^{\perp, i}} &=  D\ket{{\psi_{\bm{\theta}}}^{\perp, i}} \text{ for }  u+1 \leq i \leq d \, .
\end{align}
 Now, we calculate the eigenvalue $\lambda_1$ of the eigenstate $\ket{\lambda_1} = \hat{K} \ket{\psi_{\bm{\theta}}} / \sqrt{p_{\bm{\theta}}}$:
\begin{align}
\hat{\rho}_{\bm{\theta}} ^{\text{n, ps}} \ket{\lambda_1} & = \dfrac{1}{\PPS} \Big[ (1-\epsilon) \hat{K} \hat{\rho}_{\bm{\theta}} \hat{K}^{\dagger} + \dfrac{\epsilon}{d} \hat{K} \hat{K}^{\dagger} \Big]  \dfrac{\hat{K} \ket{\psi_{\bm{\theta}}}}{\sqrt{p_{\bm{\theta}}}} \, ,\\
    & = \dfrac{1}{\PPS} \Big[ (1-\epsilon) \hat{K} \ket{\psi_{\bm{\theta}}} \bra{\psi_{\bm{\theta}}} \hat{K}^{\dagger} \dfrac{\hat{K} \ket{\psi_{\bm{\theta}}}}{\sqrt{p_{\bm{\theta}}}} + \dfrac{\epsilon}{d} \hat{K}  \dfrac{\hat{F} \ket{\psi_{\bm{\theta}}}}{\sqrt{p_{\bm{\theta}}}} \Big] \, ,\\
    & = \dfrac{(1-\epsilon +  \dfrac{\epsilon}{d} ) \, p_{\bm{\theta}}}{\PPS}  \dfrac{\hat{K} \ket{\psi_{\bm{\theta}}}}{\sqrt{p_{\bm{\theta}}}}  = \lambda_1 \ket{\lambda_1} \, .
\end{align}
Similarly, the eigenvalues $\lambda_2,\dots,\lambda_u$ of the eigenstates $\ket{\lambda_i} = \hat{K} \ket{{\psi_{\bm{\theta}}}^{\perp, i}} / \sqrt{B}$, for $2 \leq i \leq u$, are:
\begin{align}
  \hspace{-0.25cm} \hat{\rho}_{\bm{\theta}} ^{\text{n, ps}} \ket{\lambda_i} & = \dfrac{1}{\PPS} \Big[ (1-\epsilon) \hat{K} \hat{\rho}_{\bm{\theta}} \hat{K}^{\dagger} + \dfrac{\epsilon}{d}\hat{K} \hat{K}^{\dagger} \Big] \dfrac{ \hat{K} \ket{{\psi_{\bm{\theta}}}^{\perp, i}}}{\sqrt{B}} , \\
    %& = \dfrac{1}{\PPS} \Big[ (1-\epsilon) K \ket{\psi_{\bm{\theta}}} \bra{\psi_{\bm{\theta}}} K^{\dagger} K \ket{{\psi_{\bm{\theta}}}^{\perp, i}} + \dfrac{\epsilon}{d} K  \hat{F} \ket{{\psi_{\bm{\theta}}}^{\perp, i}} \Big] \\
    & = \dfrac{\epsilon}{d \, \PPS} \hat{K} \dfrac{ \hat{F} \ket{{\psi_{\bm{\theta}}}^{\perp, i}}}{\sqrt{B}} , \\
    % & =  \dfrac{ \epsilon}{ d \, \PPS} \braket{{\psi_{\bm{\theta}}}^{\perp, m} |  \hat{F} | {\psi_{\bm{\theta}}} ^{\perp, n}}  \hat{K} \ket{{\psi_{\bm{\theta}}}^{\perp, m}} \\
    % & = \dfrac{ \epsilon }{ d \,  \PPS} \bm{\delta}_{n m} K \ket{{\psi_{\bm{\theta}}}^{\perp, m}} \\
        & = \dfrac{ \epsilon B}{ d \,  \PPS}  \dfrac{ \hat{K} \ket{{\psi_{\bm{\theta}}}^{\perp, i}}}{\sqrt{B}}  = \lambda_{i} \ket{\lambda_{i}} 	\label{Eq_line_eig} \, .
\end{align}
Doing the same calculations for the eigenvalues $\lambda_{u+1},\dots,\lambda_d$ of the eigenstates $\ket{\lambda_i} = \hat{K} \ket{{\psi_{\bm{\theta}}}^{\perp, i}} / \sqrt{D}$,  we see that $B$ in Eq. \eqref{Eq_line_eig} is replaced by $D$. Therefore,
\begin{align} \label{Eq_Eigenvalues}
    \lambda_{1} & = \dfrac{(1-\epsilon + \epsilon / d) \, p_{\bm{\theta}}}{\PPS} \,  , \\
     \lambda_{i} & = \dfrac{\epsilon \, B }{d \, \PPS} \text{ for }  2 \leq i \leq u \, ,\\
      \lambda_{i} & = \dfrac{\epsilon \, D }{d \, \PPS} \text{ for }  u+1 \leq i \leq d \, .
\end{align}
Now that we have found the eigenvalues and eigenvectors of  $\hat{\rho}_{\bm{\theta}}^{\text{n, ps}}$, we can evaluate the QFIM using Eq. \eqref{Eq_QFIM}. 
 For simplicity, in the following calculation set $a = (1-\epsilon)$ and $b = \epsilon/d$. We need to calculate $\partial_{i} \hat{\rho}_{\bm{\theta}} ^{\text{n, ps}}$: 
\begin{align}
     \partial_{i} \hat{\rho}_{\bm{\theta}}^{\text{n, ps}} & = \dfrac{a}{\PPS} \hat{K} \partial_{i} \hat{\rho}_{\bm{\theta}} \hat{K}^{\dagger}  - a\dfrac{\partial_{i} p_{\bm{\theta}}}{({\PPS})^2} \Big[ a \hat{K} \hat{\rho}_{\bm{\theta}} \hat{K}^\dag  + b  \hat{K} \hat{K}^\dag \Big]   \\
    & = \dfrac{a \, p_{\bm{\theta}}}{\PPS} (\ket{\partial_i\lambda_1} \bra{\lambda_1} + \ket{\lambda_1} \bra{\partial_i\lambda_1})   \notag \\
    & + \dfrac{a  \partial_{i} p_{\bm{\theta}}}{\PPS} \ket{\lambda_1} \bra{\lambda_1} - a^2 p_{\bm{\theta}} \dfrac{\partial_{i} p_{\bm{\theta}}}{({\PPS})^2}  \ket{\lambda_1} \bra{\lambda_1}   \\
    & = \dfrac{a \, p_{\bm{\theta}}}{\PPS} \partial_i \hat{\rho}_1 + a b \dfrac{(\partial_i p_{\bm{\theta}})}{({\PPS})^2} ( p_{\bm{\theta}} + d-1 ) \hat{\rho}_1 \label{Eq_derivative} \, ,
\end{align}
where we introduced
\begin{equation}
    \hat{\rho}_1 = \ket{\lambda_1} \bra{\lambda_1} \, .
\end{equation}
In Eq. \eqref{Eq_derivative} we also made use of the fact that
\begin{align}
    \ket{\partial_i\lambda_1} &= \left(\dfrac{\hat{K}\ket{\partial_i \psi_{\bm{\theta}}}}{\sqrt{p_{\bm{\theta}}}} - \dfrac{\hat{K}\ket{\psi_{\bm{\theta}}}}{2\sqrt{p_{\bm{\theta}}^3}}\partial_i p_{\bm{\theta}} \right)\, ,
\end{align}
so that
\begin{align}
    \dyad{\partial_i\lambda_1}{\lambda_1} + \dyad{\lambda_1}{\partial_i\lambda_1} &= \dfrac{1}{p_{\bm{\theta}}} \hat{K} \partial_{i} \hat{\rho}_{\bm{\theta}} \hat{K}^{\dagger} - \dfrac{\partial_{i} p_{\bm{\theta}}}{p_{\bm{\theta}}}\dyad{\lambda_1}\, .
\end{align}
Consider a term of the form $\matrixel{\lambda_n}{\partial_i \hat{\rho}_{\bm{\theta}}}{\lambda_m}$, as appearing in Eq. \eqref{Eq_QFIM}. If $m=n$ or both $m,n>1$, then $\matrixel{\lambda_n}{\partial_i \hat{\rho}_1}{\lambda_m}=0$ (taking the derivative of $\innerproduct{\lambda_1}\equiv 1$ implies that $\expval{\partial_i \hat{\rho}_1}{\lambda_1}=0$). On the other hand if $m\neq n$, then $\expval{\hat{\rho}_1}{\lambda_1}=0$. Further, note that, because the terms of the form $\ket{\partial_i \lambda_1}$ appearing in the QFIM are contained in the subspace $\mathcal{U}$, the only non-zero contribution comes from $m, n \leq u $. Thus Eq. \eqref{Eq_QFIM} breaks up into 2 sums: one with $n=1$ and $m\neq 1$ and one with $m=n=1$. The first of these sums is given by
% \begin{equation} 
% 2 \dfrac{a^2 p_{\bm{\theta}}^2}{\PPS ^2} \sum_{n,m}{\dfrac{ \braket{\lambda_n|  \partial_j  \hat{\rho}_1 | \lambda_m}  \braket{\lambda_m| \partial_i \hat{\rho}_1 | \lambda_n}}{\lambda_n + \lambda_m}} \, .
% \end{equation}
% We see that the term with $k = j = 1$ is zero because, in this case, the numerator is proportional to $\partial_{\bm{\theta}}{\braket{1|1}} = \braket{\dot{1} | 1 } +  \braket{1 | \dot{1}}$. Similarly, if $n, m \neq 1$, the numerator is also zero because of the orthogonality of the eigenstates of $\hat{\rho}_{\bm{\theta}}^{\text{n, ps}}$. Hence, only terms with $n = 1$ and $m \neq 1$ (or vice-versa) are non-zero:
\begin{align}
& 4 \dfrac{a^2 p_{\bm{\theta}}^2}{({\PPS})^2} \sum_{n = 2}^{u} \text{Re} \left( \dfrac{ \braket{\partial_j{\lambda_1} |\lambda_n }  \braket{\lambda_n| \partial_i{\lambda_1}}}{\lambda_n + \lambda_1}\right)\, ,  \\
& = 4 \dfrac{a^2 p_{\bm{\theta}}^2}{\PPS} \dfrac{\text{Re}( \bra{\partial_j{\lambda_1}} (\hat{1} - \ket{\lambda_1} \bra{\lambda_1})  \ket{\partial_i{\lambda_1}})}{(a + b) \, p_{\bm{\theta}} +  b B}\, , \\
& = 4 \dfrac{a^2 p_{\bm{\theta}}^2}{\PPS} \dfrac{\text{Re}( \braket{\partial_j {\lambda_1} | \partial_i{\lambda_1}} - \braket{\partial_j{\lambda_1} | \lambda_1} \braket{ \lambda_1 | \partial_i{\lambda_1} } ) }{(a + b) \, p_{\bm{\theta}} +  b B } \label{Eq_l3},\\
& = \dfrac{a^2 p_{\bm{\theta}}^2}{\PPS \Big[(a + b) \, p_{\bm{\theta}} +  b B \Big]} \mathcal{I}_{i,j} (\theta |     \hat{\rho}_{\bm{\theta}} ^{\text{ps}})   \label{Eq_l4}\, .
\end{align} 
The second term is $m=n=1$:
\begin{align}
& 2 a^2 b^2 \dfrac{ (\partial_{i} p_{\bm{\theta}}) (\partial_j p_{\bm{\theta}})}{({\PPS})^4} \Big[ p_{\bm{\theta}} + B(u-1) + D m \Big] ^2   {\dfrac{ \Big| \bra{\lambda_1} \hat{\rho}_1 \ket{\lambda_1} \Big|^2}{2\lambda_1}} \, ,\\
& = \dfrac{  a^2 b^2 \, (\partial_{i} p_{\bm{\theta}})(\partial_{j} p_{\bm{\theta}})}{({\PPS})^4} \dfrac{\Big[ p_{\bm{\theta}} + B(u-1) + D m \Big] ^2}{\lambda_1} \, , \\
& =  \dfrac{  a^2 b^2 \, (\partial_{i} p_{\bm{\theta}})(\partial_{j} p_{\bm{\theta}})}{(a+b) ({\PPS})^3 \, p_{\bm{\theta}}} \Big[ p_{\bm{\theta}} + B(u-1) \Big] ^2 \, .
\end{align}
Adding the two terms together, we find: 
\begin{equation} \label{Eq_almost_final}
     \mathcal{I}_{i,j} (\bm{\theta}  |     \hat{\rho}_{\bm{\theta}} ^{\text{n, ps}})  = \dfrac{\epsilon}{d}  \dfrac{ (1-\epsilon)^2  \, (\partial_{i} p_{\bm{\theta}})(\partial_{j} p_{\bm{\theta}}) }{\Big( 1- \epsilon + \dfrac{\epsilon}{d} \Big) \, ({\PPS})^3 \, p_{\bm{\theta}}} \Big[ p_{\bm{\theta}} +B(u-1) + D m \Big] ^2 + \dfrac{1}{\PPS} \dfrac{ (1- \epsilon)^2 p_{\bm{\theta}}^2}{  \Big(1-\epsilon + \dfrac{\epsilon}{d} \Big) \, p_{\bm{\theta}} + \dfrac{\epsilon}{d} B} \, \mathcal{I}_{i,j}(\bm{\theta} | {\hat{\rho}_{\bm{\theta}} ^{\text{ps}})}  \, .
\end{equation} 
Finally, note that
\begin{align}
    \partial_{i} p_{\bm{\theta}} &= (1-\epsilon) \Tr( \Big[ \dyad{\partial_i\psi_{\bm{\theta}}}{\psi_{\bm{\theta}}} + \dyad{\psi_{\bm{\theta}}}{\partial_i\psi_{\bm{\theta}}} \Big] \hat{K}^\dag\hat{K}),\\
    &= (1-\epsilon) t^2(\innerproduct{\psi_{\bm{\theta}}}{\partial_i\psi_{\bm{\theta}}} + \innerproduct{\partial_i\psi_{\bm{\theta}}}{\psi_{\bm{\theta}}})\, ,\\
    &= (1-\epsilon) t^2 \partial_i(\innerproduct{\psi_{\bm{\theta}}}) = 0\, ,
\end{align}
and thus the first term in Eq. \eqref{Eq_almost_final} is zero. In general, for $\bm{\delta} \neq 0$,  $p_{\bm{\theta}} = t^2 + \mathcal{O}(|\bm{\delta}|^2)$, hence  $(\partial_{i} p_{\bm{\theta}}) = \mathcal{O} (|\bm{\delta}|)$ and the lowest order correction is of order $\mathcal{O}(|\bm{\delta}|^2)$. 
Hence, we can write:
\begin{align} 
 \mathcal{I}_{i,j} (\bm{\theta}  & |     \hat{\rho}_{\bm{\theta}} ^{\text{n, ps}})  = \dfrac{1}{\PPS} \dfrac{ (1- \epsilon)^2 p_{\bm{\theta}}^2}{  \Big(1-\epsilon + \dfrac{\epsilon}{d} \Big) \, p_{\bm{\theta}} + \dfrac{\epsilon}{d} B} \, \mathcal{I}_{i,j}(\bm{\theta} | {\hat{\rho}_{\bm{\theta}} ^{\text{ps}})}  \, .
\end{align}

As in the noiseless case, the family of Filters in Eq. \eqref{Eq_filters_considered} scale of the entries of the QFIM by the same factor. 

\section{Filter Optimization}

\noindent In this Appendix, we treat noisy postselection explicitly, for some small values of $u$ and $d$. We fully optimize the filter for $u=d=2$, and then show that non-diagonal filters are advantageous for $u=2,d=3$. \\

Since most of our states are non-normalized, it is first useful to explicitly deal with normalization. Consider the case of a single parameter:
\begin{equation}
    \nr = \dfrac{\nnr}{\spps},
\end{equation}
where $\spps = \Tr(\nnr)$ normalises the state (and in our case will be the probability of passing the filter). We can then calculate 
\begin{equation}
    \partial \nr = \dfrac{\partial\nnr}{\spps} - \dfrac{\partial \spps}{(\spps)^2}\nr.
\end{equation}
Decompose the symmetric logarithmic derivative $\nsld$ as 
\begin{equation}
    \nsld = \nnsld - \dfrac{\partial \spps}{\spps}\hat{1}.
\end{equation}
It is then easy to see that $\nnsld$ satisfies the ``reduced" equation
\begin{equation}
    \dfrac{1}{2}\{\nnsld, \nnr\} = \partial \nnr.
\end{equation}
We can calculate the quantum Fisher information:
\begin{align}
    \mathcal{I}(\theta | \nr ) &= \Tr(\partial \nr \nsld ),\\
            &= \Tr\left[ \left( \dfrac{\partial\nnr}{\spps} - \dfrac{\partial \spps}{\spps}\nr \right) \left( \nnsld - \dfrac{\partial \spps}{\spps}\hat{1} \right) \right], \\
            &= \dfrac{1}{\spps}\left[\mathcal{I}(\theta | \nnr) -\dfrac{\partial \spps}{\spps} \Tr(\nnr\nnsld) - \dfrac{\partial \spps}{\spps}\Tr(\partial \nnr) +  \dfrac{(\partial \spps)^2}{\spps}\right],\\
            &= \dfrac{1}{\spps}\left[\mathcal{I}(\theta | \nnr) - \dfrac{(\Tr[\partial \nnr])^2}{\spps}\right] \label{eqn:info_rate}.
\end{align}

\subsection{Full filter optimization in $d=u=2$ (qubit sensor)} \label{App_qubit-filter}
\noindent Fix some point $\bm{\theta}$ in the parameter space. For the moment, consider variations of a single parameter; take $\partial = \partial_1$. We expand
\begin{equation}\label{eqn:deriv_expand}
    \ket{\partial \psi_{\bm{\theta}}} = ix \ket{\psi_{\bm{\theta}}} + \alpha \ket{\psi_\perp},
\end{equation}
where $\ket{\psi_{\bm{\theta}}}, \ket{\psi_\perp}$ are an orthonormal basis of our Hilbert space\footnote{We have that $\alpha\in\mathbb{C}$ may be a general complex number, but since $\innerproduct{\psi_{\bm{\theta}}}\equiv 1$, $x$ must be purely real.}. Choosing a Filter corresponds to picking a 2x2 matrix (with respect to the aforementioned basis) $\hat{K}$ such that 
\begin{equation}\label{Eq_Filter_constraint}
    0\leq \hat{K}^\dag\hat{K} \leq \hat{1}
\end{equation}
If one attempts to optimize directly over 2x2 matrices, constraint \eqref{Eq_Filter_constraint} is highly non-trivial, so we first parameterize $\hat{K}$ to simplify \eqref{Eq_Filter_constraint}.

Suppose that we have some choice of filter $\hat{K}$, we can write it in its singular value decomposition $\hat{K} = \hat{V}\hat{D}\hat{W}$, where $\hat{D}$ is diagonal with non-negative real eigenvalues, $\hat{W}$ is an SU(2) matrix and $\hat{V}$ is a U(2) matrix. Let $\hat{K}' = \hat{V}^\dag \, \hat{K}$, then the postselected state is changed by a unitary and thus the QFI and postselection probabilities do not change. Therefore $\hat{K}'$ has exactly the same performance as $\hat{K}$ and it is sufficient to consider filters of the form $\hat{D}\hat{W}$. Note that $\hat{K}^\dag\hat{K} = \hat{W}^\dag \hat{D}^2 \hat{W}$ and constraint \eqref{Eq_Filter_constraint} is equivalent to $0\leq\hat{D}\leq \hat{1}$. Using the general form of an SU(2) unitary, our filter is thus parameterised by
\begin{equation}
    \hat{D} = \begin{pmatrix}
        a &0\\
        0 & b
    \end{pmatrix}, \text{ where } 0\leq a,b \leq 1, 
    \quad
    \hat{W} = \begin{pmatrix}
        \gamma &-\beta^* \\
        \beta  &\gamma^*
    \end{pmatrix}, \text{ where } \gamma,\beta\in\mathbb{C},\, |\gamma|^2 + |\beta|^2 = 1 \, , 
\end{equation}
and $\hat{K} = \hat{D}\hat{W}$. The postselected state is then given by

\begin{equation}
    \hat{\rho}_{\bm{\theta}}^\text{n, ps} = \dfrac{ (1-\epsilon) \hat{D} \hat{W}\ket{\psi_{\theta}} \bra{\psi_{\theta}} \hat{W}^{\dagger} \hat{D}  + \dfrac{\epsilon}{2}\hat{D}^2 }{\PPS} : = \dfrac{\hat{\rho}_{\bm{\theta}}'}{\PPS} \, , 
\end{equation}
where $\PPS = \Tr(\hat{\rho}_{\bm{\theta}}')$ is the probability of postselection. 
One can calculate
\begin{align}
    \hat{\rho}_{\bm{\theta}}' &= \frac{1}{2}\begin{pmatrix}
        a^2 \epsilon + 2(1-\epsilon)|\gamma|^2 &2ab(1-\epsilon)\gamma\beta^*\\
        2ab(1-\epsilon)\gamma^*\beta & b^2\epsilon + 2(1-\epsilon)|\beta|^2 
    \end{pmatrix}\\
    \partial \hat{\rho}_{\bm{\theta}}' &= (1-\epsilon)\begin{pmatrix}
        -a^2(\alpha\gamma^*\beta^* + \alpha^*\gamma\beta) &ab(\alpha^*\gamma^2-\alpha (\beta^*)^2)\\
        ab(\alpha(\gamma^*)^2-\alpha^*\beta^2) &b^2(\alpha\gamma^*\beta^* + \alpha^*\gamma\beta)
    \end{pmatrix} \, .
\end{align}
We thus find
\begin{equation}
    \PPS = \Tr(\hat{\rho}_{\bm{\theta}}') =  \frac{1}{2}\left[(a^2+b^2)\epsilon + 2(1-\epsilon)(a^2 |\gamma|^2 + b^2 |\beta|^2)\right]\, .
\end{equation}
By finding the symmetric logarithmic derivative, we use Eq. \eqref{eqn:info_rate} to calculate

\begin{equation}
    \mathcal{I}(\bm{\theta} | \hat{\rho}_{\bm{\theta}}^\text{n, ps} ) = \frac{4a^2b^2(1-\epsilon)^2|\alpha|^2}{({\PPS})^2}\, ,
\end{equation}
Recall that we wish to maximize $\mathcal{I}(\bm{\theta} | \hat{\rho}_{\bm{\theta}}^\text{n, ps} )$, whilst minimizing $\PPS$. Consider the optimal values of $\gamma,\beta$ for fixed values of $a,b$. Since $\gamma, \beta$ only appear in $\PPS$ both of our goals are achieved by minimizing $\PPS$ subject to $|\gamma|^2+|\beta|^2=1$. This is equivalent to minimizing
\begin{equation}
    f_{a,b}(\gamma) = a^2 |\gamma|^2 + b^2 (1-|\gamma|^2).
\end{equation}
If $a\geq b$ this is clearly minimized for $|\gamma|= 0$, otherwise if $a<b$ it is clearly minimized for $|\gamma|=1$. Let
\begin{equation}
    \hat{\Omega} = \begin{pmatrix}
      0 &1\\
      -1 &0
    \end{pmatrix}.
\end{equation}
Then we can see that the best possible filter performance is attained by a filter with 
\begin{equation}
     \hat{W}=\begin{cases}
       \hat{1}_2, \quad \text{for } a < b,\\
       \hat{\Omega}, \quad \text{for } a \geq b.
     \end{cases}
\end{equation}
If $W=\hat{1}_2$ then $\hat{K}$ (and hence $\hat{F}$) are both diagonal. Otherwise note that $\hat{\Omega}^T \hat{K}$ is a diagonal filter with the same filter performance. Hence in either case the best filter performance is attained by a diagonal filter.   

Now consider the full multiparameter QFIM. Since $d=2$, we can express any state as a linear combination of $\{\ket{\psi_{\bm{\theta}}}, \ket{\psi_\perp}\}$ and thus the expansion in Eq. \eqref{eqn:deriv_expand} will hold for any parameter derivative (for different values of $x$ and $\alpha$). The optimum scaling for the diagonal terms of the QFIM is then achieved by a diagonal filter (as seen above) and thus if this filter also scales the off diagonal terms then it will be optimal. Indeed in Appendix \ref{Appendix_NoiseBefore} we show such a diagonal filter does scale the off diagonal terms, and thus the optimal filter in the multiparameter case is also diagonal.

\subsection{Non-diagonal filters are better in u=2, d=3}\label{app:off_diag_filter}
\noindent For notational brevity, we consider the case of a single parameter $\theta$. We have some filter (with Kraus operator) $\hat{K}$, a pure state $\ket{\psi_\theta}$ and a depolarizing noise rate $\epsilon$. For brevity, we will set $\epsilon=1/2$ so that it can be absorbed into the normalisation constant. We then find
\begin{equation}
    \nnr = \hat{K}\dyad{\psi_\theta}\hat{K}^\dag + \hat{K}\hat{K}^\dag/d.
\end{equation}
Fix some $\theta$, then we can decompose
\begin{equation}
    \ket{\partial \psi_\theta} = ix \ket{\psi_\theta} + \alpha \ket{\psi_\perp},
\end{equation}
where $\ket{\psi_\theta}, \ket{\psi_\perp}$ are orthonormal. Assume our Hilbert space $\mathcal{H}$ is 3-dimensional; the minimal dimension such that $\mathcal{H}$ is not spanned by $\ket{\psi_\theta}, \ket{\psi_\perp}$. Thus we can find a third normalised state $\ket{n}$ orthogonal to $\ket{\psi_\theta}$ and $\ket{\psi_\perp}$. Consider the representation of $\hat{K}$ in this basis. Take the canonical choice $(t-1)\dyad{\psi_\theta} + \hat{\Pi}_u$ from the noiseless case, with the addition of a single off-diagonal term $b\in\mathbb{R}$
\begin{equation}
    \hat{K} = \begin{pmatrix}
        t &0 &b\\
        0 &1 &0\\
        0 &0 &0
    \end{pmatrix}.
\end{equation}

We can calculate
\begin{equation}
    \hat{F} = \hat{K}^\dag \hat{K} = \begin{pmatrix}
        t^2 &0 & tb\\
        0 &1 &0\\
        tb &0 &b^2
    \end{pmatrix}.
\end{equation}

Note $\hat{F}$ has eigenvalues $0,1,t^2+b^2$. Thus as long as $0\leq b \leq \sqrt{1-t^2}$, we have $0\leq \hat{F} \leq \hat{1}$ as required. We then find
\begin{equation}
    \nnr = \begin{pmatrix}
        (4/3)t^2 &0 &tb/3\\
        0 & 1/3 & 0\\
        tb/3 & 0 & b^2/3
    \end{pmatrix}
    \quad\text{and}\quad
    \partial \nnr = \begin{pmatrix}
        0 & \alpha^* t &0\\
        \alpha t &0 &0\\
        0 &0 &0
    \end{pmatrix}.
\end{equation}
One can explicitly solve for $\nnsld$:
\begin{equation}
    \nnsld = \begin{pmatrix}
        0 &\dfrac{6(1+b^2)t \alpha^*}{1 + b^2 + (4+3b^2)t^2} &0\\
        \dfrac{6(1+b^2)t \alpha}{1 + b^2 + (4+3b^2)t^2} &0 &-\dfrac{6bt^2\alpha}{1 + b^2 + (4+3b^2)t^2}\\
        0 & -\dfrac{6bt^2\alpha^*}{1 + b^2 + (4+3b^2)t^2} &0
        \end{pmatrix}
\end{equation}

Giving a quantum fisher information of 
\begin{align}\label{eqn:off_diag_info}
    \mathcal{I}(\theta | \nnr) &= \dfrac{12|\alpha|^2(1+b^2)t^2}{1 + b^2 + (4+3b^2)t^2},\\
        &= \dfrac{12|\alpha|^2t^2}{1 + \left(3+\dfrac{1}{1+b^2}\right)t^2},\\
        &= \PPS \mathcal{I}(\theta | \nr)
\end{align}
using Eq. \eqref{eqn:info_rate} and noting $\partial \hat{\rho}_{\theta}'$ is traceless. The information rate is manifestly increasing in $|b|$ and thus it is maximized by setting $b=\sqrt{1-t^2}$, i.e. maximal mixing. 

We can compare this with the canonical choice in the noiseless case: $(t-1)\dyad{\psi_\theta}+ \hat{1}$, which is diagonal in our basis. Consider a slightly more general filter $\hat{K}'$, which we take to be diagonal:
\begin{equation}
    \hat{K}'= \begin{pmatrix}
        t & 0 &0\\
        0 & 1 &0\\
        0 &0 & r
    \end{pmatrix},
\end{equation}
where $r\in[0,1]$ controls the probability of transmitting an $\ket{n}$ state (which can only come from noise).

In this case one can similarly calculate the information rate 
\begin{equation}\label{eqn:diag_info}
    \PPS\mathcal{I}(\theta | \hat{\rho}_\theta) = \dfrac{12 |\alpha|^2 t^2}{1+4t^2},
\end{equation}
which is found to be independent of $r$. 

Comparing equations \eqref{eqn:off_diag_info} and \eqref{eqn:diag_info}, we see that any non-zero value of $b$ increases the information rate. Thus mixing is beneficial and the optimal filter in the noiseless case is not optimal in the noisy case.

\section{Optimal filter for detector saturation}\label{App_dectector_saturation_calculation}
In this Appendix we find the optimal values of $B$ and $p_{\bm{\theta}}$ for when detector saturation is the limiting factor. We recall from the main text [Eq. \eqref{Eq_TotalQFIM}] that we are attempting to solve the following optimization problem
\begin{align} \label{Eq_original_optimization}
 \text{maximise:} \quad\quad &\eta(p_{\bm{\theta}}, B) = \dfrac{ \Big( 1 - \epsilon + 2 \dfrac{\epsilon}{d} \Big)  p_{\bm{\theta}} }{ \displaystyle \Big( 1-\epsilon + \dfrac{\epsilon}{d} \Big) \frac{p_{\bm{\theta}}}{B}  + \dfrac{\epsilon}{d}} \, ,\\[10pt]
 \text{subject to:}\quad\quad & 0\leq p_{\bm{\theta}}, B \leq 1, \notag\\
        & (1-\epsilon + \frac{\epsilon}{d})p_{\bm{\theta}} + \frac{\epsilon}{d}(u-1)B\leq P_{\text{max}}.\notag
\end{align}

For notational brevity, we define $x=p_{\bm{\theta}}, y = B, b = (1-\epsilon + \frac{\epsilon}{d}), c = \frac{\epsilon}{d}(u-1), g = \frac{\epsilon}{d}$. In this notation, we see that the optimization problem \eqref{Eq_original_optimization} is equivalent to

\begin{align} \label{Eq_new_optimization}
 \text{maximise:} \quad\quad &f(x,y) = \frac{(b+g)x}{b\dfrac{x}{y}+g} \, ,\\[10pt]
 \text{subject to:}\quad\quad & 0\leq x,y \leq 1, \notag\\
        & bx + cy \leq P_{\text{max}}.\notag
\end{align}

We note that $f(x,y)$ is a strictly increasing function of $x$ and $y$ (for $x,y\geq0$) and thus the optimum point will be on the boundary of the feasible set. We deduce that if $(1,1)$ is in the feasible set it will be the optimum point. This occurs iff. $b+c\leq P_{\text{max}}$. \\

\begin{figure}[ht]
    \centering
    %% Creator: Inkscape 1.1.2 (b8e25be833, 2022-02-05), www.inkscape.org
%% Accompanies image file 'feasible_region.pdf' (pdf, eps, ps)
\begingroup%
  \makeatletter%
  \providecommand\color[2][]{%
    \errmessage{(Inkscape) Color is used for the text in Inkscape, but the package 'color.sty' is not loaded}%
    \renewcommand\color[2][]{}%
  }%
  \providecommand\transparent[1]{%
    \errmessage{(Inkscape) Transparency is used (non-zero) for the text in Inkscape, but the package 'transparent.sty' is not loaded}%
    \renewcommand\transparent[1]{}%
  }%
  \providecommand\rotatebox[2]{#2}%
  \newcommand*\fsize{\dimexpr\f@size pt\relax}%
  \newcommand*\lineheight[1]{\fontsize{\fsize}{#1\fsize}\selectfont}%
  \ifx\svgwidth\undefined%
    \setlength{\unitlength}{166.02800745bp}%
    \ifx\svgscale\undefined%
      \relax%
    \else%
      \setlength{\unitlength}{\unitlength * \real{\svgscale}}%
    \fi%
  \else%
    \setlength{\unitlength}{\svgwidth}%
  \fi%
  \global\let\svgwidth\undefined%
  \global\let\svgscale\undefined%
  \makeatother%
  \begin{picture}(1,0.79119201)%
    \lineheight{1}%
    \setlength\tabcolsep{0pt}%
    \put(0,0){\includegraphics[width=\unitlength,page=1]{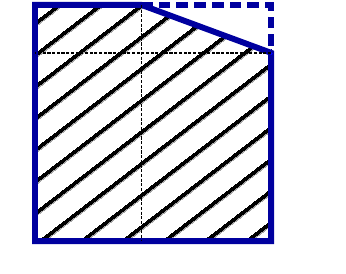}}%
    \put(0.38357723,0.01724569){\makebox(0,0)[lt]{\lineheight{1.25}\smash{\begin{tabular}[t]{l}$x_{*}$\end{tabular}}}}%
    \put(0.06025419,0.60987105){\rotatebox{90}{\makebox(0,0)[lt]{\lineheight{1.25}\smash{\begin{tabular}[t]{l}$y_{*}$\end{tabular}}}}}%
  \end{picture}%
\endgroup%

    \caption{The feasible region (shaded) of the optimization problem \eqref{Eq_new_optimization} as a subset of the unit square. The linear constraint intersects the unit square at $(1, y_*)$ and $(x_*, 1)$.}
    \label{Fig_feasible_region}
\end{figure}

Hence, we consider $b+c > P_{\text{max}}$, in which case the feasible set is the intersection of the unit square with a line of negative gradient, as depicted in Fig. \ref{Fig_feasible_region}.

\noindent Since $f$ is strictly increasing in the $x$ and $y$ directions, its maximum will be obtained on the linear constraint, i.e. it will satisfy $bx+cy=P_{\text{max}}$. As before, we introduce $t^2 = x/y$, so that
\begin{equation}
    x(t) = \dfrac{P_{\text{max}}}{b+c/t^2},\quad y(t) = \dfrac{P_{\text{max}}}{bt^2+c}.
\end{equation}

\noindent We note that $f(x(t), y(t)) = P_{\text{max}} \, \mathcal{A}(t)$, where $\mathcal{A}(t)$ is the information amplification given by Eq. \eqref{Eq_QFIMamplification_T}:

\begin{equation}
    \mathcal{A}(t) = \dfrac{ (b+g) t^2  }{(bt^2 + c)(bt^2 + g)} \, .
\end{equation}

\noindent We thus wish to maximize $\mathcal{A}(t)$ over the possible values of $t$. Let $x_{*}$ be the smallest value of $x$ attained on the line $bx+cy=P_{\text{max}}$ whilst $0\leq x,y \leq 1$. If the line intersects the unit square on the boundary $y=1$, then $x_*>0$. Otherwise, the line intersects the square on the boundary $x=0$ and $x_* = 0$. We define $y_*$ similarly. A graphical example is given in Fig. \ref{Fig_feasible_region}. We thus see that the allowed range of $t$ is given by
\begin{equation}
    x_* \leq t^2 \leq 1/y_*.
\end{equation}
Since $x_*,y_*\leq 1$, by definition $t=1$ is always allowed. \\

\noindent We find that $\mathcal{A}(t)\to 0$ as $t\to 0, \infty$ and $\mathcal{A}(t)$ has a single stationary point, which therefore must be a maximum. As before, the maximum occurs at $t=t_{\text{pp}}$, which can be bigger or smaller than 1, depending on $u,d$ and $\epsilon$. As $\mathcal{A}(t)$ is monotonic on the ranges $[0, t_{\text{pp}}]$, $[t_{\text{pp}}, \infty)$ we see that $\mathcal{A}(t)$ will be maximized by the point in the interval $[\sqrt{x_*}, \sqrt{1/y_*}]$ that is closest to $t_{\text{pp}}$. Explicitly, there are 2 cases, depending on whether $t_{\text{pp}}$ is bigger or smaller than 1:
\begin{itemize}
    \item {
        $t_{\text{pp}}\leq 1$. We have:
        \begin{equation}
            t^2 = \max[t_{\text{pp}}^2, x_*] = \max\left[t_{\text{pp}}^2, \frac{P_{\text{max}}-c}{b}\right].
        \end{equation}
    }
    \item {
        $t_{\text{pp}}\geq 1$. We have:
        \begin{equation}
            t^2 = \min[t_{\text{pp}}^2, 1/y_*] = \min\left[t_{\text{pp}}^2, \left(\frac{c}{P_{\text{max}}-b}\right)^+\right],
        \end{equation}
        where $a^+ = a$ if $a\geq 0$ and $a^+ = \infty$ if $a<0$.
    }
\end{itemize}

\begin{figure}[ht]
    \centering
    %% Creator: Inkscape 1.1.2 (b8e25be833, 2022-02-05), www.inkscape.org
%% Accompanies image file 'optimum_path.pdf' (pdf, eps, ps)
\begingroup%
  \makeatletter%
  \providecommand\color[2][]{%
    \errmessage{(Inkscape) Color is used for the text in Inkscape, but the package 'color.sty' is not loaded}%
    \renewcommand\color[2][]{}%
  }%
  \providecommand\transparent[1]{%
    \errmessage{(Inkscape) Transparency is used (non-zero) for the text in Inkscape, but the package 'transparent.sty' is not loaded}%
    \renewcommand\transparent[1]{}%
  }%
  \providecommand\rotatebox[2]{#2}%
  \newcommand*\fsize{\dimexpr\f@size pt\relax}%
  \newcommand*\lineheight[1]{\fontsize{\fsize}{#1\fsize}\selectfont}%
  \ifx\svgwidth\undefined%
    \setlength{\unitlength}{132.2801593bp}%
    \ifx\svgscale\undefined%
      \relax%
    \else%
      \setlength{\unitlength}{\unitlength * \real{\svgscale}}%
    \fi%
  \else%
    \setlength{\unitlength}{\svgwidth}%
  \fi%
  \global\let\svgwidth\undefined%
  \global\let\svgscale\undefined%
  \makeatother%
  \begin{picture}(1,1.04827574)%
    \lineheight{1}%
    \setlength\tabcolsep{0pt}%
    \put(0,0){\includegraphics[width=\unitlength,page=1]{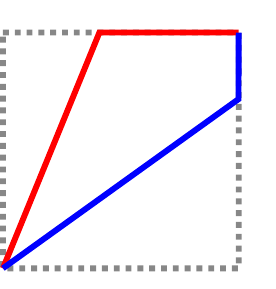}}%
    \put(0.32397711,0.98398882){\makebox(0,0)[lt]{\lineheight{1.25}\smash{\begin{tabular}[t]{l}$t_{pp}^2$\end{tabular}}}}%
    \put(0.92437348,0.71658006){\rotatebox{-90}{\makebox(0,0)[lt]{\lineheight{1.25}\smash{\begin{tabular}[t]{l}$1/t_{pp}^2$\end{tabular}}}}}%
  \end{picture}%
\endgroup%

    \caption{The path traced out by the optimum point as $P_{\text{max}}$ varies in the cases $t_{\text{pp}}\leq 1$ (red) and $t_{\text{pp}}\geq 1$ (blue). For a fixed value of $P_{\text{max}}$ the optimum value is given by the intersection of the line $bx+cy=P_{\text{max}}$ and the path of the optimum point if such an intersection exists, otherwise it is given by (1,1). }
    \label{Fig_optimum_path}
\end{figure}

\noindent This solution is best described geometrically. Consider starting $P_{\text{max}}$ at 1 and then decreasing it to zero. The maximum point of $f$ then traces out a path $(x(P_{\text{max}}),y(P_{\text{max}}))$. If $t_{\text{pp}}\leq 1$ this path moves along the boundary $y=1$ until it reaches the point $(t_{\text{pp}}^2,1)$. At this point, it moves towards the origin, along the line $x/y = t_{\text{pp}}^2$. If $t_{\text{pp}}\geq 1$ the behavior is mirrored, with the optimum point now moving along the y-axis to $(1,1/t_{\text{pp}}^2)$ before moving along the line $x/y = t_{\text{pp}}^2$ to the origin. The two types of path are sketched in Fig. \ref{Fig_optimum_path}.

\section{Noisy information amplification scaling for the JAL filter} \label{Appendix_C}
\noindent In this Appendix we show that, in the case of noisy postselection, the information amplification for the JAL filter is unchanged to first order in $\bm{\delta}$. 

Referring back to the Taylor expansion in Appendix \ref{App_JAL_Noiseless_Scaling}, we now keep only the terms to first order in $\bm{\delta}$:
\begin{equation}
    \hat{\rho}_{\bm{\theta}_0} = \hat{\rho}_{\bm{\theta}} + i  [\hat{\rho}_{\bm{\theta}} , \hat{D}] + \mathcal{O}(|\bm{\delta}|^2 ) \, , 
\end{equation}
where $\hat{D} = - i  [ \nabla_{\bm{\theta}} \hat{U}(\bm{\theta})] ^{\text{T}}\bm{\delta} U(\bm{\theta}) ^{\dagger} $. The filter $\hat{F}$ can then be written as
\begin{equation}
    \hat{F} = (t^2 -1) \hat{\rho}_{\bm{\theta}_0} + \hat{1} = (t^2 -1) \hat{\rho}_{\bm{\theta}} + \hat{1} + i [\hat{\rho}_{\bm{\theta}} , \hat{D}] (t^2 -1)  + \mathcal{O}(| \bm{\delta}| ^2 ) \, .
\end{equation}
Denote the unperturbed eigenvectors (when $\bm{\delta} = 0$) by $\ket{\lambda_1} = \hat{K} \ket{\psi_{\bm{\theta}}} / t$ and $\ket{\lambda_n} = \hat{K} \ket{{\psi_{\bm{\theta}}}^{\perp, n}}$ (for $ 2 \leq n \leq d$),with corresponding eigenvalues
\begin{equation}
    \lambda_1 = \dfrac{t^2}{\PPS} \Big( 1- \epsilon + \dfrac{\epsilon}{d} \Big) \, , \hspace{1cm}  \lambda_n = \dfrac{\epsilon }{d \PPS} \hspace{0.3cm} \text{for} \hspace{0.3cm} n \geq 2 \, , 
\end{equation}
We denote the true, perturbed eigenvectors and eigenvalues by $\ket{\lambda_n'}$,  $\lambda_n '$. In the following derivation, we will find the leading order corrections in $\bm{\delta}$ of the perturbed eigenvectors and eigenvalues. Using these corrections, we will then calculated the perturbed QFIM and show it is unchanged to order $\bm{\delta}$. 

Looking at the action of $\hat{\rho}_{\bm{\theta}}^{n, \text{ps}}$ on $\ket{\lambda_1} = \hat{K} \ket{\psi_{\bm{\theta}}} / t$, we see that
\begin{align}
    \hat{\rho}_{\bm{\theta}} ^{\text{n, ps}}  \ket{\lambda_1} &  = \dfrac{1}{\PPS} \Big[ (1-\epsilon) \hat{K} \hat{\rho}_{\bm{\theta}} \hat{K}^{\dagger} + \dfrac{\epsilon}{d}\hat{F}^\dag\Big] \hat{K} \dfrac{\ket{\psi_{\bm{\theta}}}}{t},  \\
    & =  \dfrac{1}{\PPS} \Big[ (1-\epsilon) \dfrac{\hat{K} \ket{\psi_{\bm{\theta}}}}{t} \braket{\psi_{\bm{\theta}} | \hat{F} | \psi_{\bm{\theta}}} + \dfrac{\epsilon}{d} \dfrac{\hat{K} \ket{\psi_{\bm{\theta}}}}{t} \braket{\psi_{\bm{\theta}} | \hat{F} | \psi_{\bm{\theta}}}  + \dfrac{\epsilon}{d} \sum_n \dfrac{\hat{K} \ket{{\psi_{\bm{\theta}}}^{\perp, n}}}{t} \braket{{\psi_{\bm{\theta}}}^{\perp, n} | \hat{F} | \psi_{\bm{\theta}}}   \Big], \\
    & =  \dfrac{1}{\PPS} \Big[ (1-\epsilon) t^2 \ket{\lambda_1} + \dfrac{\epsilon}{d}  t^2 \ket{\lambda_1} + i \sum_n \dfrac{\epsilon}{d \, t} (t^2 -1) \braket{{\psi_{\bm\theta}}^{\perp, n} | [\hat{\rho}_{\bm{\theta}} , \hat{D}] | \psi_{\bm{\theta}}} \ket{\lambda_n} \Big] + \mathcal{O}(\bm{\delta}^2),  \\
    & = \dfrac{t^2}{\PPS} \Big( 1- \epsilon + \dfrac{\epsilon}{d} \Big) \ket{\lambda_1} - i \dfrac{\epsilon (t^2 -1) }{d \PPS t}  \sum_{n} \braket{{\psi_{\bm{\theta}}}^{\perp, n} | \hat{D} | \psi_{\bm{\theta}}} \ket{\lambda_n} + \mathcal{O}(\bm{\delta}^2) \, .
\end{align}
Carrying out the same calculation for $\ket{\lambda_n} = \hat{K} \ket{{\psi_{\bm{\theta}}}^{\perp, n}}$
\begin{align}
    \hspace{-0.4cm} \hat{\rho}_{\bm{\theta}} ^{\text{n, ps}} \ket{\lambda_n} &  = \dfrac{1}{\PPS} \Big[ (1-\epsilon) \hat{K} \hat{\rho}_{\bm{\theta}} \hat{K}^{\dagger} + \dfrac{\epsilon}{d}\hat{F}^\dag\Big] \hat{K} \ket{{\psi_{\bm{\theta}}}^{\perp, n}}, \\
    & =  \dfrac{1}{\PPS} \Big[ (1-\epsilon) \dfrac{\hat{K} \ket{\psi_{\bm{\theta}}}}{t} t \braket{\psi_{\bm{\theta}} | \hat{F} | {\psi_{\bm{\theta}}}^{\perp, n} } + \dfrac{\epsilon}{d} \dfrac{\hat{K} \ket{\psi_{\bm{\theta}}}}{t} 	t  \braket{\psi_{\bm{\theta}} | \hat{F} | {\psi_{\bm{\theta}}}^{\perp, n} }  + \dfrac{\epsilon}{d} \sum_{m} \hat{K} \ket{{\psi_{\bm{\theta}}}^{\perp, m}}  \braket{{\psi_{\bm{\theta}}}^{\perp, m} | \hat{F} | {\psi_{\bm{\theta}}}^{\perp, n}}  \Big],  \\
    & = i \dfrac{t}{\PPS}  \Big( 1- \epsilon + \dfrac{\epsilon}{d} \Big) (t^2 -1) \braket{  \psi_{\bm{\theta}} | \hat{D} | {\psi_{\bm{\theta}}}^{\perp, n} } \ket{\lambda_1} +  \dfrac{\epsilon }{d \PPS}  \ket{\lambda_n} + \mathcal{O}(\bm{\delta}^2) \, .
\end{align}
In the above calculations, we used the following identities: 

\begin{align}
 \braket{\psi_{\bm{\theta}} |  [\hat{\rho}_{\bm{\theta}}, \hat{D}]| \psi_{\bm{\theta}}} 
    & =  \braket{\psi_{\bm{\theta}} | \hat{\rho}_{\bm{\theta}} \hat{D}  - \hat{D} \hat{\rho}_{\bm{\theta}}| \psi_{\bm{\theta}}}  \\
    & =  \braket{\psi_{\bm{\theta}} | \hat{D} | \psi_{\bm{\theta}}} - i \braket{\psi_{\bm{\theta}} | \hat{D} | \psi_{\bm{\theta}}}  =0 \, ,
\end{align}
\begin{align}
 \braket{{\psi_{\bm{\theta}}}^{\perp, n}  |  [\hat{\rho}_{\bm{\theta}}, D]| \psi_{\bm{\theta}}} 
    & =  \braket{{\psi_{\bm{\theta}}}^{\perp, n}  | \hat{\rho}_{\bm{\theta}} \hat{D}  - \hat{D} \hat{\rho}_{\bm{\theta}}| \psi_{\bm{\theta}}} \\
    & = -  \braket{{\psi_{\bm{\theta}}}^{\perp, n}  | \hat{D} | \psi_{\bm{\theta}}} \, ,
\end{align}
\begin{align}
 \braket{{\psi_{\bm{\theta}}}^{\perp, n}  |  [\hat{\rho}_{\bm{\theta}}, \hat{D}]|{\psi_{\bm{\theta}}}^{\perp, m}} 
    & =  \braket{{\psi_{\bm{\theta}}}^{\perp, n}  | \hat{\rho}_{\bm{\theta}} \hat{D}  - \hat{D} \hat{\rho}_{\bm{\theta}}| {\psi_{\bm{\theta}}}^{\perp, m}} = 0 \, .
\end{align}
Therefore, $\hat{\rho}_{\bm{\theta}} ^{\text{n, ps}}$ can be written in the $\{\ket{\lambda_1}, \dots, \ket{\lambda_d} \}$ basis as
\begin{equation}
    \hspace{-1cm} \hat{\rho}_{\bm{\theta}} ^{\text{n, ps}} = 
    \begin{pmatrix}
     \dfrac{t^2}{\PPS} \Big( 1- \epsilon + \dfrac{\epsilon}{d} \Big) & i \dfrac{t}{\PPS}  \Big( 1- \epsilon + \dfrac{\epsilon}{d} \Big) (t^2 -1) \braket{  \psi_{\bm{\theta}} | \hat{D} | {\psi_{\bm{\theta}}}^{\perp, n} }  &  \dots & i \dfrac{t}{\PPS}  \Big( 1- \epsilon + \dfrac{\epsilon}{d} \Big) (t^2 -1) \braket{  \psi_{\bm{\theta}} | \hat{D} | {\psi_{\bm{\theta}}}^{\perp, n} }  \\
       - i \dfrac{\epsilon (t^2 -1) }{d \PPS t}  \braket{{\psi_{\bm{\theta}}}^{\perp, n} | \hat{D} | \psi_{\bm{\theta}}}   & \dfrac{\epsilon }{d \PPS} & \dots & 0 \\
        \vdots & \vdots &  \ddots & \vdots \\
            - i \dfrac{\epsilon (t^2 -1) }{d \PPS t}  \braket{{\psi_{\bm{\theta}}}^{\perp, n} | \hat{D} | \psi_{\bm{\theta}}}  & 0 & \dots & \dfrac{\epsilon }{d \PPS} \\
    \end{pmatrix} \, .
\end{equation}
We find the perturbed eigenvalues by solving $\text{det} ( \hat{\rho}_{\bm{\theta}} ^{\text{n, ps}} - \lambda \, \hat{1}) = 0$, which is equivalent to
\begin{equation}
    \Big[ \dfrac{t^2}{\PPS} \Big( 1- \epsilon + \dfrac{\epsilon}{d} \Big) - \lambda \Big] \cdot \Big[ \dfrac{\epsilon }{d \PPS} - \lambda \Big]^{d-1} = \Big[ \dfrac{\epsilon (t^2 -1) }{d \PPS t}  \braket{{\psi_{\bm{\theta}}}^{\perp, n} | \hat{D} | \psi_{\bm{\theta}}} \Big] \cdot \Big[ \dfrac{t}{\PPS}  \Big( 1- \epsilon + \dfrac{\epsilon}{d} \Big) (t^2 -1) \braket{  \psi_{\bm{\theta}} | \hat{D} | {\psi_{\bm{\theta}}}^{\perp, n} }  \Big] \cdot \Big[ \dfrac{\epsilon }{d \PPS} - \lambda \Big]^{d-2}  (d-1) \, .
\end{equation}
Hence, we have
\begin{equation}
    \Big[ \dfrac{\epsilon }{d \PPS} - \lambda \Big]^{d-2} \Bigg\{ \Big[ \dfrac{t^2}{\PPS} \Big( 1- \epsilon + \dfrac{\epsilon}{d} \Big) - \lambda \Big] \Big[ \dfrac{\epsilon }{d \PPS} - \lambda \Big] - \dfrac{\epsilon (d-1)}{d ({\PPS})^2} \Big( 1- \epsilon + \dfrac{\epsilon}{d} \Big)  (t^2 -1)^2 |\braket{  \psi_{\bm{\theta}} | \hat{D} | {\psi_{\bm{\theta}}}^{\perp, n} }  |^2 \Bigg\}  = 0 \, .
\end{equation}
Therefore, the eigenvalues are the same up to a term of order $\bm{\delta}^2$, i.e. $\lambda_{n} ' = \lambda_{n} + \mathcal{O}(\bm{\delta}^2)$.

From the matrix form of $ \hat{\rho}_{\bm{\theta}} ^{\text{n, ps}}$, we see that the perturbed eigenvectors will have corrections of order $\bm{\delta}$
\begin{gather}
    \ket{\lambda_1 '} = \ket{\lambda_1} +  \sum_{n \neq 1 } \alpha_{1, n} \ket{\lambda_n} + \mathcal{O}(\bm{\delta}^2)  \, , \\
    \ket{\lambda_n '} = \ket{\lambda_n} +  \alpha_n \ket{\lambda_1}  + \mathcal{O}(\bm{\delta}^2) \hspace{0.3cm} \text{for} \hspace{0.3cm} n > 2   \, ,
\end{gather}
where $\alpha_i$ are terms of order $\bm{\delta}$. 

We can now evaluate the first order correction in $\bm{\delta}$ of the QFIM. We use the formula  
\begin{equation} 
2 \dfrac{a^2 p_{\bm{\theta}}^2}{({\PPS})^2} \sum_{n,m}{\dfrac{ \braket{\lambda_n ' | \big( \partial_i  \hat{\rho}_1   \big) | \lambda_m '}  \braket{\lambda_m '|\big( \partial_j \hat{\rho}_1  \big) | \lambda_n '}}{\lambda_n '  + \lambda_m '  }} \, ,
\end{equation}
where $\hat{\rho}_1  = \ket{\lambda_1 }\bra{\lambda_1 }$ and $\lambda_n '$, $\ket{\lambda_n '}$ are now the perturbed eigenvalues and eigenvectors. To evaluate $\hat{\rho}_{\bm{\theta}} ^{\text{n, ps}}$ we used Eq. \eqref{Eq_derivative} and neglected the term proportional to $\partial_i p_{\bm{\theta}}$, because it is second order in $\bm{\delta}$. We also used the fact that $p_{\bm{\theta}} = t^2 + \mathcal{O}(\bm{\delta}^2)$. Note that the eigenvalues $\lambda_n'$ are unchanged to order $\bm{\delta}$. Let us examine the following cases:

\begin{itemize}
    \item $n = m = 1$: the first factor in the numerator is proportional to $ \sum_{l \neq 1} \alpha_{1,l}^* \braket{\lambda_l | \partial_i \lambda_1} + \sum_{l \neq 1} \alpha_{1,l} \braket{ \partial_i \lambda_1 | \lambda_l} $, which is order $\bm{\delta}$. Indeed, notice that the first-order term cancels out: $\braket{\lambda_1 | \partial_i \lambda_1} + \braket{ \partial_i \lambda_1 | \lambda_1} = \partial_i \braket{\lambda_1 | \lambda_1 } = 0$. When multiplied by the second factor (the $j$-derivative), this gives a second-order contribution in $\bm{\delta}$.
    \item $n, m \neq 1$: the first term in the numerator is proportional to $ \alpha_{m}^* \braket{\lambda_n | \partial_i \lambda_1} +  \alpha_{n} \braket{ \partial_i \lambda_1 | \lambda_m} $, which is first order in $\bm{\delta}$. Similarly, the second term is also first order in $\bm{\delta}$. The contribution from these terms is second-order in  $\bm{\delta}$. 
    \item $n \neq 1 , m = 1$ (or vice-versa): the first factor in the numerator is given by 
    \begin{align}
& \braket{\lambda_n | \partial_i \lambda_1}  + \alpha_n ^* \braket{\lambda_1 | \partial_i \lambda_1}  + \alpha_n ^* \braket{\partial_i \lambda_1 |  \lambda_1}  +  \mathcal{O}(|\bm{\delta}|^2)  \\
& = \braket{\lambda_n | \partial_i \lambda_1} +  \mathcal{O}(|\bm{\delta}|^2) \, , \label{Eq_F1}
\end{align}
where we used $\braket{\lambda_1 | \partial_i \lambda_1} + \braket{ \partial_i \lambda_1 | \lambda_1} = \partial_i \braket{\lambda_1 | \lambda_1 } = 0$. Similarly, to order $\bm{\delta}^2$, the second factor in the numerator is $\braket{\partial_j \lambda_1 | \lambda_n}$. When this term is multiplied by the first term in  Eq. \eqref{Eq_F1}, we recover the formula for the unperturbed QFIM calculated in the main text, to order $\bm{\delta}^2$. 
\end{itemize}

\end{document}